\documentclass[12pt,preprint]{aastex}  

\usepackage{natbib}
\usepackage{epsfig}
\usepackage{psfig}

\newcommand{\be}{\begin{equation}}
\newcommand{\ee}{\end{equation}}
\newcommand{\ba}{\begin{eqnarray}}
\newcommand{\ea}{\end{eqnarray}}
\newcommand{\bas}{\begin{eqnarray*}}
\newcommand{\eas}{\end{eqnarray*}}

\newcounter{saveeqn}

\begin{document}

\title{Improved Cosmological Constraints from Gravitational Lens Statistics}

\author{Jonathan L. Mitchell,${}^{1,2}$ 
Charles R. Keeton,${}^{1,}$\altaffilmark{3,4}
Joshua A. Frieman,${}^{1,2,5}$
\& Ravi K. Sheth${}^{6}$}

\affil{${}^1$Dept.\ of Astronomy and Astrophysics, University of Chicago, 
5640 South Ellis Avenue, Chicago, IL 60637}
\affil{${}^2$Center for Cosmological Physics, University of Chicago,
5640 South Ellis Avenue, Chicago, IL 60637}
\affil{${}^5$NASA/Fermilab Astrophysics Center, Fermi National Accelerator 
Laboratory, P.O.\ Box 500, Batavia, IL 60510}
\affil{${}^6$Dept.\ of Physics and Astronomy, University of Pittsburgh,
Pittsburgh, PA 15620}
\altaffiltext{3}{Hubble Fellow.}
\altaffiltext{4}{Present address: Department of Physics and Astronomy,
Rutgers University, 136 Frelinghuysen Road, Piscataway, NJ 08854}

\begin{abstract}
We combine the Cosmic Lens All-Sky Survey (CLASS) with new
Sloan Digital Sky Survey (SDSS) data on the local velocity
dispersion distribution function of E/S0 galaxies, $\phi(\sigma)$, 
to derive lens statistics constraints on $\Omega_\Lambda$ and
$\Omega_m$.
Previous studies of this kind relied on a combination of the
E/S0 galaxy luminosity function and the Faber-Jackson relation
to characterize the lens galaxy population.
However, ignoring dispersion in the Faber-Jackson relation 
leads to a biased estimate of $\phi(\sigma)$ and therefore 
biased and overconfident constraints on the cosmological
parameters.  
The measured velocity dispersion function from a large sample
of E/S0 galaxies provides a more reliable method for probing
cosmology with strong lens statistics. 
Our new constraints are in good agreement with recent results
from the redshift-magnitude relation of Type Ia supernovae.
Adopting the traditional assumption that the E/S0 velocity
function is constant in comoving units, we find a maximum
likelihood estimate of $\Omega_\Lambda = 0.74$--$0.78$ for
a spatially flat unvierse (where the range reflects uncertainty
in the number of E/S0 lenses in the CLASS sample), and a 95\%
confidence upper bound of $\Omega_\Lambda<0.86$.
If $\phi(\sigma)$ instead evolves in accord with extended
Press-Schechter theory, then the maximum likelihood estimate
for $\Omega_\Lambda$ becomes $0.72$--$0.78$, with the 95\%
confidence upper bound $\Omega_\Lambda<0.89$.  Even without
assuming flatness, lensing provides independent confirmation 
of the evidence from Type Ia supernovae for a nonzero dark 
energy component in the universe.
\end{abstract}


\section{Introduction}

Gravitationally lensed quasars and radio sources offer important
probes of cosmology and the structure of galaxies.  The optical
depth for lensing depends on the cosmological volume element out
to moderately high redshift, so lens statistics can in principle
provide valuable constraints on the cosmological constant or, more
generally, the dark energy density and its equation of state 
\citep[e.g.,][]{ffk90, fukugita and turner, turner, krauss and white,
maoz and rix, kochanek96, falco, cooray, waga, wagaf, srm, chae et al., 
chae}.

However, the cosmological constraints derived from lens statistics  
have been controversial, mainly because of disagreements about the
population of galaxies that can act as deflectors.
Kochanek (1996; see also Falco et al.\ 1998; Kochanek et al.\ 1998)
reported an upper bound of $\Omega_\Lambda < 0.66$ at 95\% confidence
for a spatially flat universe ($\Omega_m + \Omega_\Lambda = 1$),
which is in marginal conflict with the current concordance
cosmology, $\Omega_\Lambda = 0.69 \pm 0.04$ \citep{spergel}.
But subsequent studies have reached different conclusions
\citep[e.g.,][]{chiba,waga,cheng}.
For example, \citet{chiba} argued that optically-selected lenses
actually favor $\Omega_\Lambda = 0.7^{+0.1}_{-0.2}$ for a flat
universe. 
At issue are uncertainties in several key ingredients of traditional
lens statistics calculations:
(i) the luminosity function for early-type (E/S0) galaxies, which
dominate the lensing rate;
(ii) the Faber-Jackson relation between luminosity and velocity
dispersion for early-types;
and (iii) the assumed density profiles of lens galaxies.
The spread in derived cosmological constraints can be traced in
large measure to uncertainties in the galaxy luminosity function:
until recently, different redshift surveys yielded values for the
local density of $L^*$ galaxies that differed by up to a factor of
two.  This source of uncertainty has now been largely eliminated by
much larger galaxy redshift surveys, such as 
the Sloan Digital Sky Survey (SDSS) and
the 2dF Galaxy Redshift Survey (2dFGRS)
\citep{blanton1,blanton2,yasuda,norberg,madgwick}. 

Even with the local galaxy luminosity function well determined, 
there is a crucial systematic uncertainty concerning changes to
the deflector population with redshift.  Many analyses of lens
statistics have assumed that the velocity dispersion distribution
function $\phi(\sigma)$ is independent of redshift (in comoving
units).  This is equivalent to saying that massive
early-type\footnote{Late-type galaxies only constitute a small
fraction of the lensing optical depth, so evolution in that
population is not very important for lens statistics.} galaxies
have not undergone significant mergers since $z \sim 1$.  Although
galaxy counts appear to be consistent with this `no-evolution' model
in the concordance cosmology \citep{schade,im}, the observational
uncertainties are still large and other possibilities cannot be
ruled out.  The problem for lens statistics is that evolution is
degenerate with cosmology.  \citet{keeton02a} has argued that
previous studies obtained strong limits on $\Omega_\Lambda$ only
because they assumed that the evolution rate is independent of
cosmology;\footnote{The fact that they assumed the evolution rate
to be zero is actually less important than the fact that they
assumed it to be independent of cosmology; see \citet{keeton02a}.}
dropping that assumption would make lens statistics largely
insensitive to cosmology.  One way to handle the degeneracy is
to turn the problem around: adopt values for the cosmological
parameters and attempt to constrain models of galaxy evolution
\citep[e.g.,][]{ofek,chae-evol}.  Unfortunately, the small size of
current samples precludes using more than toy models of evolution,
and even then the uncertainties are too large to distinguish a
simple no-evolution model from various theoretical predictions.
It would still be nice to use lens statistics to probe cosmology
while accounting for evolution using more than toy models.

The problems with the traditional approach to lens statistics
have partly motivated an alternate approach, in which empirical
calibrations of the deflector population are replaced with
theoretical predictions from galaxy formation models
\citep[e.g.,][]{narayan, kochanek95, porciani, KM01, srm, li}.
In these theory-based models, the deflector population is
described by a dark matter halo mass function, $n(M,z)$, given
by Press-Schechter theory
\citep[calibrated by N-body simulations, see][]{st,smt,jenkins}.
The predicted mass function depends on cosmology, which causes
the lensing optical depth to depend on $\Omega_\Lambda$ through
the cosmological volume element, the density perturbation growth
rate, and the merger histories of halos.
Unlike in the traditional approach, here the optical depth
{\it decreases\/} with increasing $\Omega_\Lambda$ --- suggesting
that the traditional lensing upper bound on $\Omega_\Lambda$
should be interpreted with caution. 
This `theoretical' approach to lens statistics avoids some of
the untested assumptions of the traditional approach and has the
advantage of working directly with the deflector mass function
rather than indirectly with a mass function inferred from the
galaxy luminosity function. 
However, it faces challenges of its own, chiefly arising from
theoretical uncertainties in relating dark matter halos to the
properties of luminous galaxies. 
For example, galaxy formation models have difficulty reproducing
the observed galaxy luminosity function and empirical galaxy
dynamical scaling relations \citep[e.g.,][]{wf91, cole94,
kauffmann93, kauffmann99, somerville, benson03}.
Since nearly all confirmed lens systems contain a luminous galaxy
that plays a significant role in the lensing, the problems with
galaxy formation models may cause concern about the theoretical
approach to lens statistics.

The goal of this paper is to make two modifications to lens
statistics calculations that enable improved cosmological constraints.
The first modification involves using new data on the dynamical
properties of galaxies.  In standard models, the lensing optical
depth is given by a weighted integral over the galaxy velocity
dispersion distribution function, $\phi(\sigma)$
\citep[][also see \S 2.2]{TOG}.  Previously, $\phi(\sigma)$ was
inferred by combining the measured early-type galaxy luminosity
function $\phi(L)$ with the empirical Faber-Jackson relation,
$L(\sigma)$; hereafter, we call this the {\it inferred\/} velocity
function. This estimator for $\phi(\sigma)$ has two disadvantages: 
(i) neglect of the scatter in the Faber-Jackson relation yields a
biased and incorrectly confident estimate for $\phi(\sigma)$
\citep{kochanek94, sheth et al.}; and
(ii) use of the luminosity function complicates attempts to deal
with galaxy evolution, since $\phi(L)$ is sensitive not only to
dynamical galaxy number and mass evolution (which matter for lens
statistics) but also to passive luminosity evolution (which does
not affect lens statistics).
To obviate these problems, it is preferable to use a direct
measurement of the E/S0 velocity function.  Fortunately, the SDSS
recently provided this very measurement based on $\sim$30,000 E/S0
galaxies \citep{bernardi,bernardi2,sheth et al.}.
With these new data, we can eliminate an important source of bias
and misestimated error in lens statistics calculations.

The second modification concerns galaxy evolution.  To make contact
with previous studies, we consider models in which $\phi(\sigma)$
is constant in comoving units.  However, we also study models in
which $\phi(\sigma)$ evolves according to a theoretical prescription.
As just mentioned, the fact that $\phi(\sigma)$ evolves only due to
occasional mergers means that it provides a more straightforward
framework for incorporating evolution than the traditional route
through the luminosity function.  Newman \& Davis (2000, 2002)
present such a framework using extended Press-Schechter theory
to compute the ratio of the velocity function at redshift $z$ to
the local velocity function, $\phi(\sigma;z)/\phi(\sigma;0)$.
While model predictions for the full velocity function $\phi(\sigma,z)$
are sensitive to the uncertain physics that causes discrepancies
between galaxy formation models and observed galaxy populations,
the prediction for the ratio $\phi(\sigma,z)/\phi(\sigma,0)$
 isolates the evolution piece \citep{newman2} and is therefore
much less sensitive to these uncertainties.
By joining the theoretical evolution model to the empirical
calibration of the local deflector population, we obtain a new
hybrid approach to lens statistics that combines the best aspects
(and omits the pitfalls) of the purely empirical or purely theoretical
approaches used previously.  In the end, we find that inclusion of the 
extended Press-Schechter model for evolution does not significantly 
change the central values of the cosmological parameters inferred from 
lensing statistics, though it does increase the associated uncertainties.

For the lens sample, we use the Cosmic Lens All Sky Survey 
\citep[CLASS;][]{CLASS,browne}, which is the largest statistically
complete survey for lenses.  \citet{chae et al.} and \citet{chae}
recently analyzed the CLASS sample using the traditional approach
based on an inferred velocity function.  We use the same sample
but analyze it using our new approach to lens statistics.  Other
small technical differences between the analyses are discussed
below.

The layout of the paper is as follows.  In \S 2 we review the 
theoretical framework, including lensing by isothermal spheres,
the formalism for lens statistics, and the model for redshift 
evolution of the deflector population.  In \S 3 we discuss the
required observational data, including the measured and inferred
velocity dispersion distribution functions from the SDSS early-type
galaxy sample, and the CLASS radio lens survey.  In \S 4 we use a
likelihood analysis of the lens data to derive constraints on
cosmological parameters.  We conclude in \S 5.  In the Appendix
we discuss the SDSS early-type galaxy selection process and its 
effect on our model inputs.

\section{Theoretical Framework}

\subsection{The singular isothermal sphere lens}

X-ray studies \citep[e.g.,][]{fabbiano},
dynamical analyses \citep[e.g.,][]{rix,gerhard},
and various lensing studies \citep[e.g.,][]{treu,koopmans,rkk}
all indicate that on the $\lesssim$10~kpc scales relevant for
lensing, early-type galaxies can be modeled as singular isothermal
spheres (SIS), with a density profile corresponding to a flat
rotation curve,
\be \rho (r)= \frac{\sigma^2}{2\pi Gr^2}\ . \ee
Here $\sigma$ is the velocity dispersion of the system, $r$ is
the distance from the center of the galaxy, and we have assumed
negligible core radii and ellipticities.  While lens statistics
are in principle sensitive to finite-density cores in lens galaxies
\citep[e.g.,][]{chiba,cheng,hinshaw}, the elusiveness of
`core images' limits the sizes of cores to a level that is
unimportant \citep{krauss and white, wallington, rusin and ma, 
keeton02b, winn}.
Also, while departures from spherical symmetry are important in
detailed models of individual lenses \citep[e.g.,][]{kks}, they
have remarkably little effect on lens statistics.  \citet*{huterer}
show any biases from neglecting ellipticity and shear in lens
statistics analyses are at the level of $\Delta\Omega_M<0.01$ and
$\Delta\Omega_\Lambda<0.02$ (also see \S 4.5 below).

Consider light rays propagating from a source past a lens to the
observer.  For an SIS lens with velocity dispersion $\sigma$, 
the ray bending angle is $4\pi (\sigma/c)^2$, independent of
impact parameter.  Multiple imaging occurs if the physical
impact parameter is less than $4\pi (\sigma/c)^2 (D_{OL}D_{LS})/D_{OS}$, 
where  $D_{OL}$, $D_{LS}$, and $D_{OS}$ are the angular diameter
distances from observer to lens, lens to source, and observer to
source, respectively.  It is therefore useful to define the angular
Einstein radius, 
\be \theta_E =4\pi \left(\frac{\sigma}{c}\right)^2 \frac{D_{LS}}{D_{OS}} ~, \ee
such that sources located at angle $\theta_S < \theta_E$ from an
SIS lens are multiply imaged.  For a Friedmann-Robertson-Walker
cosmology with cosmological constant $\Omega_\Lambda$,
non-relativistic matter density $\Omega_m$, and 
curvature density $\Omega_k = 1 - \Omega_\Lambda - \Omega_m$,  
the angular diameter distance can be written 
\be D_{xy} = \frac{r_{xy}}{1+z_y} =  
  \frac{c}{H_0} \frac{S_k (\chi_{xy})}
{1+z_y} \ee
where $r_{xy}$ is the transverse comoving distance, $H_0$ is the Hubble 
constant,  
\be S_k (\chi_{xy}) = \left\{ \begin{array}{ll}
      \frac{1}{\sqrt{|\Omega_{k}|}} \ \sin 
(\sqrt{|\Omega_{k}|} \ \chi_{xy}) & \mbox{if $\Omega_k<0$} \\
      \frac{1}{\sqrt{\Omega_{k}}} \ \sinh 
(\sqrt{\Omega_{k}} \ \chi_{xy}) & \mbox{if $\Omega_k>0$} \\
      \chi_{xy}  & \mbox{if $\Omega_k=0$}
      \end{array} 
    \right.   \mbox{,} \ee
and
\be \chi_{xy} = \int_{z_x}^{z_y} dz\ \left[\Omega_k(1+z)^2 + \Omega_\Lambda + 
\Omega_m(1+z)^3 \right]^{-1/2} \ . \ee
Here and throughout, we specialize to the case where the dark energy
is identical to a cosmological constant; the generalization to
a different dark energy equation of state is straightforward
\citep{waga,cooray}.  

A source at angular separation $\theta_S < \theta_E$ from an SIS lens
yields two images on opposite sides of the lens at angular positions
\be \theta_{\pm} = \theta_E \pm \theta_S \ , \ee  
which have magnifications
\be \mu_{\pm} = \frac{\theta_E \pm \theta_S}{\theta_S} ~. \ee
The image at $\theta_{-}$ has $\mu_{-} < 0$ indicating that this
image is parity reversed.  The angular separation between the
images is $\Delta\theta = 2\theta_E$, independent of the
source position.  The total magnification of the two images is
\be \mu_{tot} = \frac{2\theta_E}{\theta_S} \ , \ee
and the bright-to-faint image flux ratio is
\be f = \frac {\theta_E+\theta_S}{\theta_E-\theta_S} \ . \ee
In general lens surveys have a limited dynamic range, so a lens
will be identified only if the flux ratio is less than some
value; the CLASS survey included an explicit cut at $f_{max}=10$
(see \S 3.2).  Thus only sources with
$\theta_S < \theta_{max} < \theta_E$ will lead to detectable lenses,
where
\be \frac {\theta_{max}}{\theta_E} = \frac{f_{max}-1}{f_{max}+1} \ . \ee

\subsection{Lens statistics}

The optical depth for lensing is obtained by summing the cross
sections for all deflectors between observer and source.  Since the
SIS cross section depends only on the lens velocity dispersion and
cosmological distances, the property of the deflector population
that is directly relevant is the velocity function, $\phi(\sigma)$.
The optical depth for lensing can be written as an integral over
$\phi(\sigma)$ \citep[see, e.g.,][]{TOG},
\be \tau(z_S, \Omega_m, \Omega_\Lambda) = \frac{1}{4\pi} \int_{0}^{z_S}{dV} 
\int_{0}^{\infty}{d\sigma\ \phi(\sigma;z_L )}\
A(\sigma , \Omega_m , \Omega_\Lambda, z_L, z_S)B(S_\nu) ~, 
\label{eq:tau}
\ee
where $z_S$ and $z_L$ are the source and lens redshifts, 
$A$ is the cross section for multiple imaging, $B$ is the
magnification bias (defined below),
and the differential comoving volume element is 
\be dV = 4\pi r_{OL}^2\,
\frac{dr_{OL}}{dz_L}d\,z_L ~. \ee
For an SIS lens, the angular separation between the two images 
is always twice the Einstein radius, so we can replace the integral
over velocity dispersion with one over image separation.

Magnification bias accounts for the fact that intrinsically faint
sources can appear in a flux-limited survey by virtue of the
lensing magnification.  The product of the cross section $A$ and
the magnification bias $B$ can be written as
\be
A(\sigma,\Omega_m , \Omega_\Lambda , z_L , z_S ) B(S_\nu)
  = 2 \pi \int^{\theta_{max}}_{0} d\theta_S\ \theta_S\
  \frac{N(>\!S_0/\mu_{tot})}{N(>\!S_0)}\ ,
\label{eq:bias1}
\ee
where $N(>\!S)$ is the number of sources brighter than flux $S$,
$S_0$ is the flux limit of the survey, and it is appropriate to
use the total magnification $\mu_{tot}$ when the sources in the
original flux-limited catalog are unresolved.  If the source
counts can be modeled as a power law, $dN/dS \propto S^{-\eta}$ 
\citep[a good approximation for CLASS sources;][]{chae et al.},
then eqn.~\ref{eq:bias1} can be evaluated to be
\be
A B = \pi \theta_E^2 \times \frac{2^{\eta}}{3-\eta} 
  \left(\frac{f_{max}-1}{f_{max}+1}\right)^{3-\eta} ,
\label{eq:bias2}
\ee
for an SIS lens population.
Note that, absent a flux ratio cut, the cross section for an SIS lens
would just be $A = \pi \theta_E^2$.  It is convenient to define
a combined correction factor ${\tilde B}$ that accounts for both
magnification bias and the flux ratio limit of the lens survey, 
\be {\tilde B}(\eta, f_{max}) \equiv  \frac{A B}{\pi \theta_E^2} ~. \ee

From the total optical depth $\tau$ we can determine several
interesting statistical distributions.
$d\tau/d\Delta\theta$ describes the distribution of image
separations, 
$d\tau/d z_L$ gives the redshift distribution of lens
galaxies, and 
$d^2\tau/dz_L d\Delta\theta$ gives the joint distribution
for both the lens galaxy redshift $z_L$ and the image separation
$\Delta\theta$.
All three of these distributions, together with the total optical
depth, are used in the likelihood analysis of the CLASS survey
(see \S 4).

\subsection{A model for redshift evolution of the lens population}

Many previous studies of lens statistics have assumed the velocity
function $\phi(\sigma)$ to be constant in comoving units.  This
no-evolution assumption is usually justified by appealing to results
from galaxy number counts \citep{im,schade} and the redshift
distribution of lens galaxies \citep{ofek}, which are consistent
with the hypothesis that the early-type population evolves only
through passive luminosity evolution.  However, the observational
status of early-type evolution has been controversial
\citep{CNOC,KCW96,TY98,fried}, and the observational uncertainties
are large enough that dynamical number or mass evolution in the
early-type galaxy population cannot be ruled out.

Evolution of $\phi(\sigma)$ in amplitude or shape could
substantially impact cosmological constraints from lens statistics.
In order to gauge these effects, we adopt an evolution model
based on theoretical galaxy formation models.  Following
Newman \& Davis (2000, 2002), we use extended Press-Schechter
theory to compute the ratio of the velocity dispersion function at
two epochs, $\phi(\sigma;z)/\phi(\sigma;0)$, as a function of
cosmological parameters.  This ratio can be combined with the
measured local velocity dispersion function $\phi(\sigma;0)$ to
estimate $\phi(\sigma;z)$ at any epoch.  As discussed in the
Introduction, this estimate represents a hybrid approach to lens
statistics that combines a careful measurement of the local
velocity function with a simple but robust theoretical prediction
for evolution.

N-body simulations of structure formation in cold dark matter models 
\citep[e.g.,][]{jenkins} indicate that the halo mass function at
epoch $z$ is well fit by the modified Press-Schechter form introduced
by \citet{st}, 
\be
n(M;z) = {\bar{\rho} \over M}\,{d \ln \nu \over dM}\,A(p)
\left[1+(q\nu)^{-p}\right] \left({q \nu \over 2 \pi}\right)^{1/2}\,
\exp(-q\nu/2)\, , \label{eq:st}
\ee
where $\bar{\rho}$ is the mean density, 
$\nu(z) = \delta_c^2/\sigma^2_\delta(M,z)$, $\delta_c = 1.686$ is the 
extrapolated linear overdensity of a spherical top hat perturbation 
at the time it collapses, 
$\sigma^2_\delta(M,z)$ is the variance of the density field at epoch
$z$ in linear perturbation theory, smoothed with a top hat filter of 
radius $R=(3M/4\pi \bar{\rho})^{1/3}$, and the fitting parameters 
have values $p=0.3$, $A(p)=0.3222$, and $q=0.75$.  The smoothed 
variance is given in terms of the present linear density power 
spectrum $P(k)$ by
\be
\sigma^2_\delta(M,z) = 
{D^2(z)\over 2 \pi^2} \int_0^\infty k^2\, P(k)\, W^2(k;M)\, dk\, ,
\ee
where $W(k;M)$ is the Fourier transform of the top hat window 
function of radius $R(M)$. The linear growth factor is given by
$D(z) = I(z)/I(0)$, where 
\be
I(z) = \int_z^\infty {{1+z}\over E(z)^3}\, dz\, ,
\ee
and $E(z) = H(z)/H_0 = \left[\Omega_m(1+z)^3 + \Omega_\Lambda 
+ \Omega_k(1+z)^2\right]^{1/2}$. 

To convert the mass function into a velocity function, we must
take into account the formation epoch of halos: those that form
earlier will be more concentrated and have higher velocity
dispersion for fixed mass.  Following a simplified version of
the procedure in \citet{newman}, we use the results of \citet{lacey} 
to estimate the mean formation redshift $z_f$ for a halo of mass 
$M$ observed at redshift $z$.  \citet{lacey} define a scaled 
variable
\be
{\tilde \omega_f} = \delta_c\ { {D^{-1}(z_f) - D^{-1}(z)} \over 
[\sigma^2_\delta(M/2,0)-\sigma^2_\delta(M,0)]^{1/2} }\ ,
\label{eq:omegaf}
\ee
and the distribution of formation redshifts is given implicitly
by the probability distribution $dp/d{\tilde \omega_f}$. 
N-body simulations indicate that $dp/d{\tilde \omega_f}$ is 
nearly independent of halo mass and of the power spectrum shape 
\citep{lacey}; following their Figure~12, we approximate this 
distribution by a delta function at
$\langle {\tilde \omega_f} \rangle = 0.9$.
While this effectively ignores the dispersion of formation epoch,
we have checked that this approximation does not significantly 
affect the estimate of $\phi(\sigma;z)/\phi(\sigma;0)$ over the
range of interest. 

\begin{figure}[t]
\begin{center}
\epsfig{file=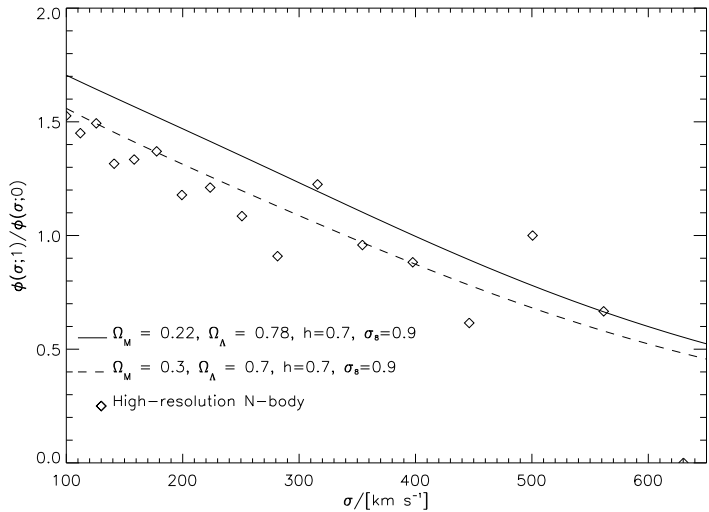,angle=0,width=4.truein}
\end{center}
\caption{The ratio of the velocity function of halos at $z=1$ to
that at $z=0$. The points show results from a high-resolution
N-body simulation by A. Kravtsov (private communication),
and the dashed curve shows results from the extended
Press-Schechter theory for the same cosmological parameters.
The solid curve shows the extended Press-Schechter prediction
for the best-fit flat cosmology from lens statistics for the
CLASS sample (see \S 4), keeping $h=0.7$ and
$\sigma_8 =0.9$ fixed.
}
\label{fig:ratio}
\end{figure}

Solving eqn.~\ref{eq:omegaf} for $z_f$, and modeling each halo as
an SIS, we can infer the velocity dispersion \citep{newman,bryan}, 
\be
\sigma(M,z)= 92.3\, \Delta_{\rm vir}(z_f)^{1/6}\, E(z_f)^{1/3}\,
\left({M \over 10^{13}\,h^{-1} M_\odot}\right)^{1/3}\
{{\rm km}\over {\rm sec}}\ ,
\label{eq:sigMz}
\ee
where \citep{bryan}
\be
\Delta_{\rm vir}(z) = 18\pi^2 + 60[\Omega(z)-1] - 32[\Omega(z)-1]^2 ~,
\ee
and
\be
\Omega(z) = {\Omega_m (1+z)^3 \over E^2(z)}~.
\ee
Combining eqn.~\ref{eq:sigMz} with eqn.~\ref{eq:st} yields the
velocity function $\phi(\sigma;z)$.  Figure~\ref{fig:ratio}
shows the ratio $\phi(\sigma;1)/\phi(\sigma;0)$ versus $\sigma$
for several sample cases.  In general, $\phi(\sigma;z)$ grows
with redshift for $\sigma$ less than a few hundred km/sec, and
the growth is strongly dependent on cosmological parameters.  We
have checked that the model agrees well with high-resolution
N-body simulations (A.~Kravstov, private communication; see
Fig.~\ref{fig:ratio}). 

In Fig.~\ref{fig:ratio}, we have also included the evolution ratio that results from the parameterized evolution by \citet{chae-evol}.  By comparing our model with \citet{chae-evol} in Fig.~\ref{fig:ratio}, one might wonder if our dynamical evolution model is missing some evolution at high $\sigma$.  However, the \citet{chae-evol} model actually captures both dynamical evolution and passive luminosity evolution, whereas our model neatly isolates the dynamical evolution.   

We caution that the Press-Schechter model describes
the behavior of dark halos, and we are assuming that it
applies to luminous, early-type galaxies.
While this ignores subtleties associated with baryonic
infall and luminous galaxy formation, the N-body simulations
shown above do have sufficiently high resolution to resolve
galactic-scale subhalos, and they are included in the
results for $\phi(\sigma)$. Moreover, SPH and semi-analytic
models indicate that moderately massive halos contain
a luminous, central galaxy. To the extent that the
measured velocity dispersions of early-type galaxies
provide good estimates of the velocity dispersions of
the sub-halos they occupy (see \S 3.1.1), this model should
provide a reasonable approximation to the evolution
of $\phi(\sigma)$ for early-type galaxies. In addition,
since we are computing a ratio (which is
generally less than 2 for the redshifts and velocity
dispersions of interest for lensing), it should not be
extremely sensitive to these subtle
effects (for details, see \citealt{newman2}).

\section{Observational Inputs}

\subsection{The deflector population}

We follow the traditional approach to lens statistics and
assume that all lenses are associated with optically luminous
galaxies and calibrate the deflector population empirically.
Furthermore, we focus on early-type galaxies.  Although late-type
galaxies are more abundant than early-types, they tend to have
lower masses and hence to contribute no more than 10--20\% of the
lensing optical depth.  This is a standard prediction of lens
statistics models \citep{TOG, fukugita and turner, maoz and rix}
that has been borne out by the data
\citep[e.g.,][]{fassnacht,kkf,kochanek00,lubin}.
We could attempt to model both the early- and late-type deflector
populations in order to compute the total lensing optical depth
and compare to the observed number of lenses produced
by early- and late-type galaxies \citep[as done by][]{chae et al.,
chae}.  However, we believe it is simpler and more instructive to
separate the galaxy types, to compute the optical depth due to
early-type galaxies alone, and to compare that to the number of
lenses produced by early-type galaxies.  This allows us to avoid
dealing with uncertainties in the description of the late-type
galaxy population.

In the following sections we describe two models for the distribution
of the early-type deflectors.  First, we use a direct measurement
of the early-type velocity dispersion function.
We then specify an inferred velocity dispersion function using
the early-type luminosity function transformed by the mean Faber-Jackson 
relation in order to make contact with and compare to previous studies of 
this type.  

\subsubsection{The measured velocity function}

We calibrate the E/S0 deflector population using a sample of 
$\sim$30,000 early-type galaxies at redshifts $0.01 \leq z \leq 0.3$
selected from the SDSS database following \cite{bernardi,bernardi2}.
A detailed description of the selection procedure is given in the 
Appendix.  Briefly, the selection is based on both morphological and 
spectral criteria:  the sample is restricted to galaxies with 
de Vaucouleurs surface brightness profiles that lack strong emission 
lines, for which measurements of the velocity dispersion are available.  

The SDSS data reduction pipelines only measure velocity dispersions 
for galaxies with spectra of sufficiently high signal to noise to 
ensure accurate measurement.  In addition, the resolution of the SDSS 
spectrographs prevents accurate estimates of dispersions smaller than 
$\sigma=70$ km/sec \citep{bernardi}.  
Since this cutoff corresponds to a typical lens image separation
of $\Delta \theta \lesssim 0\farcs14$, well below the $0\farcs3$
resolution limit of the CLASS survey, it has a negligible effect
on our analysis.  The Appendix describes various tests of the 
selection procedure which suggest that the sample does not miss more 
than, and probably much less than, 30\% of the early-type population.
We therefore disagree with the claim by \citet{chae} that the 
Bernardi et al.\ sample is too restrictive to be representative of
the early-type population, at least as regards the velocity function 
relevant for lensing.  

The SDSS E/S0 sample size has increased from the $\sim$9000 used by
\citet{bernardi} to $\sim$30,000 used by \citet{bernardi2},
purely because a larger fraction of the sky has now been observed.  There 
are small differences in the data that arise from modifications to the
SDSS data reduction pipeline; see \citet{bernardi2} for details.
Briefly, the new model magnitudes (which are used to fit the $L(\sigma)$
relation) are fainter by $\sim$0.12 mag, and the half-light radii
$\theta_{\rm eff}$ are smaller by $\sim$10\%.  The change in size causes
a small change in the velocity dispersions: while the measured dispersions
are the same, the aperture correction from the SDSS fiber radius
($\theta_{\rm fiber} = 1\farcs5$) to a uniform physical radius 
(conventionally taken to be $\theta_{\rm eff}/8$) has changed.  (It is
the aperture-corrected `central' velocity dispersions that we need,
because these are very nearly equal to the dark matter velocity
dispersions needed for the lensing calculations; see \citealt{franx, 
kochanek93,kochanek94, treu and koopmans}).  Observed velocity 
dispersion profiles
typically fall as weak power laws, so the correction has the form
$\sigma_{\rm ap-cor} = \sigma_{\rm fiber} (8 \theta_{\rm fiber}/
\theta_{\rm eff})^{0.04}$, and the decrease in $\theta_{\rm eff}$ leads to
a slight increase in the aperture-corrected velocity dispersions.  These
revisions to the SDSS photometry have affected the luminosity and
velocity functions of the E/S0 sample as well as the slope of the
$\sigma(L)$ relation.

\begin{figure}[t]
\begin{center}
\epsfig{file=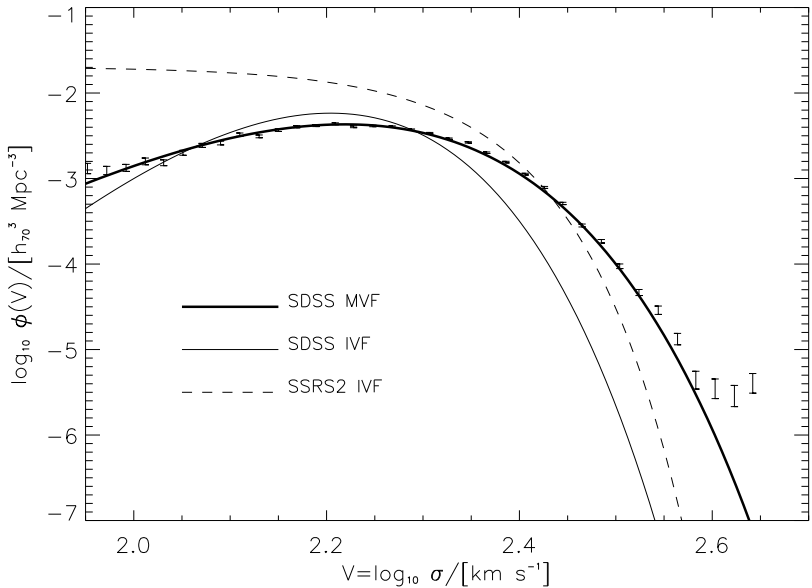,angle=0,width=4.7truein}
\end{center}
\caption{
The data points show the measured velocity function (MVF) for the
sample of $\sim$30,000 early-type galaxies in the SDSS.  The heavy
and light solid curves show the best Schechter-like fits to the
SDSS measured (MVF) and inferred (IVF) velocity functions,
respectively.  For comparison, the dashed curve shows the IVF for 
the SSRS2 early-type galaxy sample \citep{marzke}, after the
normalization correction applied by \citet{chae}.
}
\label{fig:VF}
\end{figure}

\citet{sheth et al.} use the aperture-corrected dispersions to
compute the velocity function, which is shown by the points in
Figure~\ref{fig:VF} \citep[for the revised sample from][]{bernardi2}.
The function can be modeled as a modified Schechter function
\citep{schechter} of the form
\be \phi(\sigma)\,d\sigma = \phi_* \left(\frac{\sigma}{\sigma_*}\right)^\alpha\
\exp\left[-\left(\frac{\sigma}{\sigma_*}\right)^\beta\right]\
\frac{\beta}{\Gamma(\alpha/\beta)}\ \frac{d\sigma}{\sigma}\ ,
\label{eq:schechter}
\ee
where $\phi_*$ is the integrated number density of galaxies,
$\sigma_*$ is a characteristic velocity dispersion,\footnote{Note
that $\sigma_*$ can be quite different from the mean value:
$\langle\sigma\rangle =
\sigma_*\,\Gamma[(1+\alpha)/\beta]/\Gamma[\alpha/\beta] = 160$~km/s.}
$\alpha$ is the low-velocity power-law index, and $\beta$ is the
high-velocity exponential cutoff index of the distribution.  The
best-fit parameter values are\footnote{These values are the same as
those reported by \citet{sheth et al.} for the original sample of
\citet{bernardi}, except that the normalization $\phi_*$ is lower
by 30\%.}
\be (\phi_*,\sigma_*,\alpha,\beta)_{MVF} = ((1.4\pm 0.1)\times 10^{-3} 
\ {h_{70}^{3}\rm Mpc}^{-3}, \ 88.8\pm 17.7 \ {\rm km/sec}, \ 
6.5 \pm 1.0, \ 1.93\pm 0.22) \ ,   
\label{eq:MVF}
\ee
where the Hubble parameter $H_0 = 70\,h_{70}$~km/sec/Mpc.  The
curve in Figure~\ref{fig:VF} shows this fit.  Possible evolution
in the velocity function can be treated as redshift dependence in
the parameters $\phi_*$, $\sigma_*$, $\alpha$, and/or $\beta$.

The new, larger sample of $\sim$30,000 early-type galaxies in the SDSS 
contains a small surplus of galaxies with velocity dispersions $\ge$450~km/sec
that is not fit by the Schechter function (see Fig.~\ref{fig:VF}).
Although massive, such galaxies are sufficently rare that they
contribute only $\sim$0.2\% of the lensing optical depth, so we
have not attempted to modify the MVF fit to include them.

Using the Schechter-like fit for the velocity function, the optical 
depth becomes (see eqn.~\ref{eq:tau})
  \be \tau(z_S, \Omega_m , \Omega_\Lambda ) = 
\int_{0}^{z_S} \tau_*(z_L) \left(\frac{r_{OL}r_{LS}} {r_{OS}}\right)^2
\frac{dr_{OL}}{dz_L}\ dz_L\ {\tilde B}(\eta, f_{max})
\label{eq:tauMVF} \ee
where
  \be \tau_*(z) = 16\pi^3\, \phi_*(z) \left[\frac{\sigma_*(z)}{c} \right]^4\
\frac{\Gamma\left[\frac{\alpha(z)+4}{\beta(z)}\right]}
{\Gamma\left[\frac{\alpha(z)}{\beta(z)}\right]}\ .
\label{eq:taustar}
\ee 
If there is no evolution in $\phi(\sigma)$ then $\tau_*$ is just a
constant that can be pulled out of the integral in eqn.~\ref{eq:tauMVF}.
For a flat cosmology, the redshift integral in eqn.~\ref{eq:tauMVF} 
can be evaluated analytically; in this no-evolution flat case, the 
optical depth is $\tau = \tau_*\,{\tilde B}\,(c/H_0)^3\,r_{OS}^3/{30}$.  
This simple example illustrates how lens statistics probe the volume
of the universe out to the redshifts of the sources.

\subsubsection{The inferred velocity function}

As discussed in the Introduction, previous analyses of lens
statistics usually obtained an estimate of the velocity function
by taking an observed galaxy luminosity function and transforming
it using the Faber-Jackson relation; we refer to this estimate as
the inferred velocity function, or IVF.  Generally, the luminosity
function is modeled as a Schechter function,
\be \phi(L)\,dL = {\tilde\phi}_* \left(\frac{L}{L_*}\right)^{\tilde\alpha}
\exp\left[-\left(\frac{L}{L_*}\right)\right]\ \frac{dL}{L_*}\ ,
\label{eq:oldLschechter}
\ee
where the parameters are the comoving number density of galaxies
${\tilde\phi}_{*,LF}$, the characteristic luminosity $L_*$ (or
corresponding absolute magnitude $M^*$), and the faint-end slope
$\tilde\alpha_{LF}$.  With a Faber-Jackson relation of the form
$L/L_* = (\sigma/\sigma_*)^\gamma$, the IVF becomes
\be \phi(\sigma)\,d\sigma = {\tilde\phi}_*
\left(\frac{\sigma}{\sigma_*}\right)^{\gamma({\tilde\alpha}+1)-1}\
\exp\left[-\left(\frac{\sigma}{\sigma_*}\right)^\gamma\right]\
\gamma\ \frac{d\sigma}{\sigma_*}\ .
\label{eq:oldVschechter}
\ee
The coefficient of the optical depth, $\tau_*$, for this distribution 
differs slightly from the form of eqn.~\ref{eq:taustar}:
\be \tau_* = 16\pi^3\, {\tilde\phi_*} \left(\frac{\sigma_*}{c} \right)^4\
\Gamma\left[1+{\tilde\alpha}+\frac{4}{\gamma}\right]\ .
\label{eq:lftstar}
\ee 
With this change, the optical depth has the same form as
eqn.~\ref{eq:tauMVF}.

We must consider how evolution in the deflector population could
affect the velocity function.  Dynamical evolution due to mergers
would change both the luminosity function and the velocity function.
Passive luminosity evolution (due to aging stellar populations) would
affect the luminosity function {\it but not the velocity function},
at least for simple models.  If galaxies of different luminosities
have the same passive evolution rate, then $L$ depends on redshift
but $L/L_*$ does not.  Conceptually, the changes in the luminosity
function are offset by corresponding changes in the Faber-Jackson
relation such that the IVF remains constant.  This makes sense,
because the velocity function describes the dynamical properties
of galaxies so any evolution that leaves the dynamics unchanged
must also leave the velocity function unchanged.  We focus on a
non-evolving velocity function when using the IVF.

\citet{chae} and \citet{chae et al.} recently analyzed the
statistics of CLASS lenses using an IVF based on the
Second Southern Sky Redshift Survey (SSRS2).
SSRS2 is a relatively shallow ($z \le 0.05$),
bright ($m_B \le 15.5$) survey that contained only 5404 galaxies
but allowed visual classification of the morphological types
\citep{marzke}, yielding 1595 early-type galaxies.  With this
small sample, the normalization is sure to suffer biases from
large-scale inhomogeneities; to compensate, Chae corrected the
normalization using the total luminosity function normalization
scaled by the fraction of early-types measured in other, larger
surveys.  The Schechter luminosity function parameters for the
early-type galaxy sample, as reported by \citet{chae}, are
\be ({\tilde\phi}_*,M^*_0 - 5 \ log_{10} \ h_{70},{\tilde\alpha})_{LF, \ SSRS2} 
 = (2.2\times 10^{-3} \ {h_{70}^{3}\rm Mpc}^{-3}, -20.40, \ -1.0) \ .  
\label{eq:SSRS2LF}
\ee
{\citet{chae} and \citet{chae et al.} fixed the Faber-Jackson index
at $\gamma = 4$.  Rather than using external constraints on
$\sigma_*$, they chose to calibrate this parameter as part of
their likelihood analysis of CLASS lenses.  In effect, $\sigma_*$
was determined by the distribution of lens image separations.  The
resulting best-fit IVF parameters for SSRS2 are
\be({\tilde\phi}_*,\sigma_*,\gamma{\tilde\alpha}+\gamma-1,\gamma)_
{IVF, \ SSRS2} 
= (2.2\times 10^{-3} \ {h_{70}^{3}\rm Mpc}^{-3}, \ 198 \ {\rm km/sec}, \ 
-1.0, \ 4.0) \ .
\label{eq:SSRS2IVF}
\ee 
This fit is shown as the dashed line in Figure \ref{fig:VF}.
There are two possible causes for concern in the use of the lens
image separation distribution for an internal calibration of
$\sigma_*$.  First, this approach introduces Poisson errors
associated with the small lens sample.  Second, it may introduce
systematic biases if the small number of lens galaxies in the sample
are not representative of massive early-type galaxies.  Use of the
velocity dispersion function measured directly from a large sample
avoids both of these problems.

We can also obtain an IVF for the SDSS early-type galaxy sample.
The error bars in Figure~\ref{fig:lf} show the measured luminosity
function for the revised SDSS sample from \citet{bernardi2}.  The
dashed line shows a Gaussian fit to the data reported by
\citet{bernardi}, but shifted faintwards by 0.125~mag and downwards
to $\phi_*=0.001$, as required by the new data reductions.  We
have refit the sample with a modified Schechter function
(eqn.~\ref{eq:schechter}), finding best-fit parameters
\be (\phi_*,M^*_r,\alpha,\beta)_{LF,SDSS} = (1.4\times 10^{-3} \ 
{h_{70}^{3}\rm Mpc}^{-3}, \ -16.46 - 0.85z, \ 2.53, \ 0.43) \ .
\label{eq:LF}
\ee
The solid curve in Figure~\ref{fig:lf} show this fit.  Compared to
the Gaussian fit, the Schechter fit does a better job at both the
faint end (which is why its normalization $\phi_*$ is slightly
larger) and the bright end, so we focus on it.

\begin{figure}[t]
\begin{center}
\epsfig{file=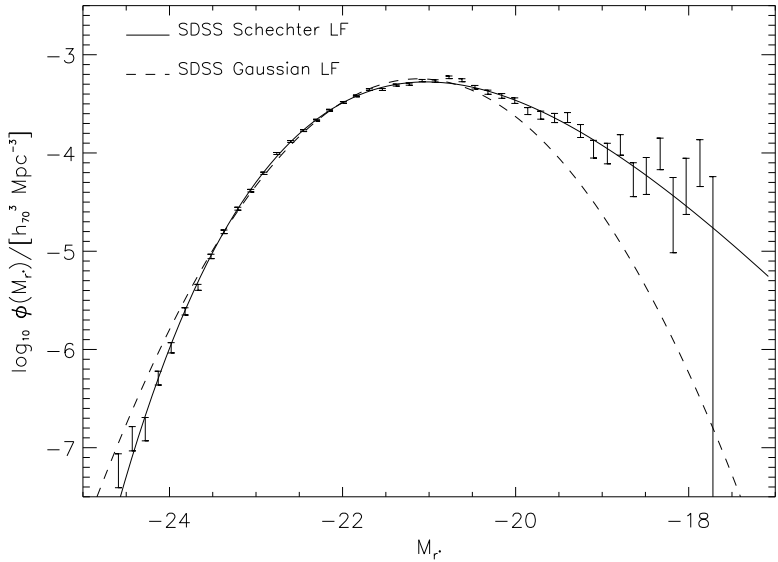,angle=0,width=4.5truein}
\end{center}
\caption{The SDSS E/S0 luminosity function.  The points show the data
from the sample of $\sim$30,000 galaxies given by \citet{bernardi2}.
The dashed line shows the Gaussian fit reported by \citet{bernardi},
with $M^*$ increased by 0.125 and the normalization reduced to
$\phi_*=0.001$ to adjust to the updated photometry
\citep[see][]{bernardi2}.  The solid line shows our Schechter fit.}
\label{fig:lf}
\end{figure}

The SDSS sample also provides a direct calibration of the
$L(\sigma)$ (Faber-Jackson) relation.  With the sample from
\citet{bernardi2}, the mean inverse relation is (see
Fig.~\ref{fig:fj} in the Appendix)
\be \langle \log_{10}(\sigma/{\rm km~sec}^{-1}) \mid M_r \rangle =
2.2 - 0.091(M_{r} + 20.79 + 0.85z)\, , 
\label{eq:ifj}
\ee
which corresponds to a Faber-Jackson index $\gamma = 4.4$.  Thus,
the SDSS IVF is described by the parameters
\be (\phi_*,\sigma_*,\alpha,\beta)_{IVF,SDSS} = (1.4\times 10^{-3} 
\ {h_{70}^{3}\rm Mpc}^{-3}, \ 64.0 \ {\rm km/sec}, \ 11.13, \ 1.89) \ ,
\ee
which is also shown in Figure~\ref{fig:VF}.

Clearly both the SSRS2 and SDSS IVFs differ systematically from the
SDSS MVF. \citet{sheth et al.} showed that the difference between
the SDSS IVF and MVF is due to the fact that the IVF ignores the 
considerable dispersion in the $L(\sigma)$ relation.  They found
that the RMS scatter around the mean inverse relation (eqn.~\ref{eq:ifj})
is
\be {\rm rms}(\log_{10} (\sigma /{\rm km~sec}^{-1})\mid M_{r}) 
= 0.79 \ [1+0.17 \ (M_r + 21.025 + 0.85z)] \ .
\label{eq:ifjscatt}
\ee
(This result holds for both the original and revised SDSS samples.)
The scatter broadens the velocity function and, in particular, 
raises the tail to high $\sigma$ without changing the mean 
\citep[also see][]{kochanek94}.  The impact on lens statistics is 
apparent when we examine the differential `lensing efficiency' (LE), 
or the contribution to the lensing optical depth from each 
$\sigma$ bin (see eqn.~\ref{eq:taustar}):
\be LE \equiv \phi(\sigma)\,\sigma^4 \propto \frac{d\tau}{d\sigma}\ . \ee
Figure~\ref{fig:LE} shows the lensing efficiency for the SSRS2 IVF,
the SDSS IVF, and the SDSS MVF.  The IVF substantially underestimates
the abundance of massive early-type galaxies and hence the total
optical depth.  This effect leads directly to a lensing estimate
for $\Omega_\Lambda$ that is biased high (see \S 5).  The effect
can be seen quantitatively by comparing
$\tau_*=6.92\times10^{-3}$ for the SDSS MVF, versus
$\tau_*=5.79\times10^{-3}$ for the SDSS IVF.  

\begin{figure}[t]
\begin{center}
\epsfig{file=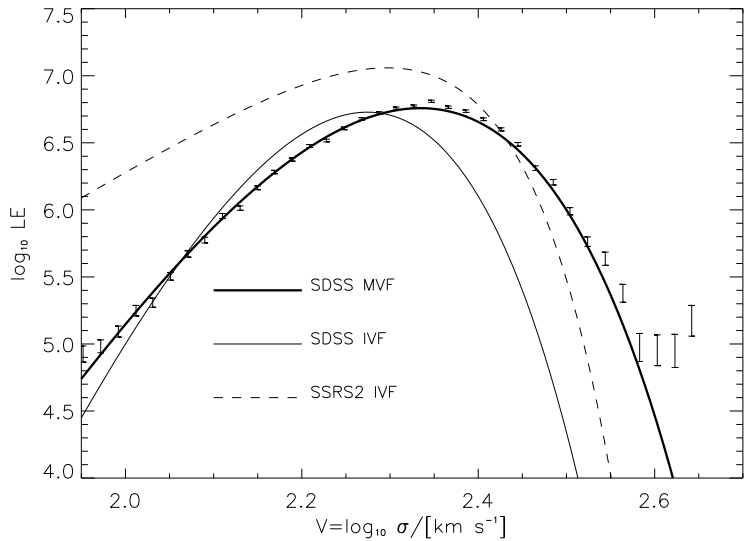,angle=0,width=4.3truein}
\end{center}
\caption{A comparison of the lensing efficiency, 
$LE \equiv \phi(\sigma)\,\sigma^4$, for the measured and inferred 
velocity functions from the SDSS early-type galaxy sample, and for
the inferred velocity function from the SSRS2 early-type galaxy
sample.}
\label{fig:LE}
\end{figure}

\subsection{Radio source lens survey: CLASS}

\begin{table}[t]
\begin{center}
\begin{tabular}{ l l c c c l l}
\hline \hline
Survey & Name &  $z_L$ & $z_S$ & $\Delta \theta$ & Lens & Reference \\
\hline 
JVAS    & B0218+357 & 0.68      & 0.96 & 0.33 & S    & \citet{B0218_357} \\
CLASS   & B0445+123 & 0.56      & ---  & 1.33 & ?    & \citet{B0445_123} \\
CLASS   & B0631+519 & ---       & ---  & 1.16 & ?    & \citet{browne} \\
CLASS   & B0712+472 & 0.41      & 1.34 & 1.27 & E    & \citet{0712disc} \\
CLASS   & B0850+054 & 0.59      & ---  & 0.68 & ?    & \citet{B0850_054} \\
CLASS   & B1152+199 & 0.44      & 1.01 & 1.56 & ?    & \citet{1152/1359disc} \\
CLASS   & B1359+154 & ---       & 3.21 & 1.65 & ?, m & \citet{1152/1359disc} \\
JVAS    & B1422+231 & 0.34      & 3.62 & 1.28 & E    & \citet{B1422_231} \\
CLASS   & B1608+656 & 0.64      & 1.39 & 2.08 & E, m & \citet{CLASS} \\
CLASS   & B1933+503 & 0.76      & 2.62 & 1.17 & E    & \citet{1933disc} \\
CLASS   & B2045+265 & 0.87      & 1.28 & 1.86 & ?    & \citet{B2045_265} \\
JVAS    & B2114+022 & 0.32/0.59 & ---  & 2.57 & E, m & \citet{B2114_022} \\
CLASS   & B2319+051 & 0.62/0.59 & ---  & 1.36 & E    & \citet{B2319_051} \\
\hline\hline
\end{tabular}
\end{center}
\caption{Data for the 13 Lenses in the CLASS statistical sample of
8958 objects (adapted from \citet{browne}, \citet{chae}, and
\citet{davis}).  ``Lens'' stands for the morphology of the lens
galaxy: spiral (S), elliptical (E), or unknown (?); three lenses
contain multiple galaxies (m).}
\label{table1}
\end{table}

While some 80 multiply imaged quasars and radio sources have been 
discovered, a statistical analysis requires a sample from a survey
that is complete and has well-characterized, homogeneous selection
criteria.  The largest such sample comes from the radio Cosmic Lens
All-Sky Survey \citep[CLASS;][]{browne,myers}, an extension of the
earlier Jodrell Bank/Very Large Array Astrometric Survey
\citep[JVAS;][]{patnaik.jvas,king}.  About 16,000 sources have been
imaged by JVAS/CLASS, with 22 confirmed lenses.  Of these, a subset
of 8958 sources with 13 lenses forms a well-defined subsample suitable
for statistical analysis \citep{browne}.  The properties of these 
lenses are summarized in Table~1.  Of the 13 lenses, 8 have measured
source redshifts, 11 have measured lens redshifts, and 7 have both 
\citep{chae et al.}.  

Radio lens surveys \citep{quast, helbig, chae et al.} have several
advantages over earlier optical QSO lens surveys:
(i) they contain more sources and therefore have smaller statistical 
errors;
(ii) they are not afflicted by systematic errors due to reddening
and obscuration by dust in the lens galaxies;
and (iii) they can more easily probe sub-arcsecond image angular
separations than seeing-limited optical surveys.
The main limitations of radio surveys is poor knowledge of the
radio source luminosity function \citep{marlow,munoz} and redshifts.

The flux limit of the CLASS survey is 30~mJy at 5 GHz.  The flux
distribution of sources above the flux limit is well described
by a power law, $|dN/dS_\nu| \propto S_\nu^{-\eta}$, with
$\eta=2.07\pm 0.02$.  The statistical lens sample is believed to
be complete for all lenses for which the flux ratio between the
images is $\le$10.  Using these parameters with eqn.~\ref{eq:bias2},
we find that the factor ${\tilde B}$ in the optical depth that
accounts for the magnification bias and the flux ratio cut is
${\tilde B} = 3.97$.

As discussed in \S 3.1, we compute the optical depth due to
early-type galaxies and seek to compare that with the number of
lenses produced by early-type galaxies in the CLASS survey.  
However, the morphologies and spectral types of the lens galaxies
have been identified in only some of the CLASS lenses: of the 13
lenses in Table~1, six are known to be E/S0 galaxies, one is a
spiral, and the rest are unknown.  With 80--90\% of lenses produced
by E/S0 galaxies, we would expect 10--12 of the CLASS lenses to
have early-type lens galaxies.  We exclude from our analysis the
one lens identified as a spiral, B0218+357.  There are arguments
for discarding two others as well: B1359+154 because it has three
lensing galaxies and our analysis cannot include the effects of
compound lenses (since the distribution of lens environments is not
known; see \S 4.5); and B0850+054 because \citet{mckean et al.}
identify its spectrum as Sb and its sub-arcsecond image separation 
might be taken to suggest that it is produced by a spiral galaxy.  
We carry out our analysis for two cases, one with 12 lenses and the 
other with 10, and we believe that this spans the plausible range of
possibilities.

In order to understand whether our MVF model correctly represents
the CLASS lens sample, we should evaluate whether the
lens galaxies in the CLASS sample would meet the criteria for the
SDSS E/S0 sample.  \citet{bernardi,bernardi2} defined their
sample using spectral and morphological cuts (see \S 3.1.1).  Due
to the high redshift of lens galaxies, however, Hubble Space Telescope (HST)
imaging is required, and not always sufficient, to perform luminosity
profile fits.  Four CLASS lenses are well fit by a de Vaucouleurs
profile.  Two of these are compound lenses: B1359+154 is a group of
three galaxies all with smooth de Vaucouleurs profiles \citep{B1359}; 
and B1608+656 is a pair of galaxies with one heavy-dust spiral and one
smooth de Vaucouleurs E/S0 \citep{B1608pro}.  In only one lens, 
B1933+503, were exponential and de Vaucouleurs fits compared, and the
de Vaucouleurs model proved better \citep{1933disc}.  
The fourth lens, B0712+472, has a concentration expected for a Sa
galaxy, but the inner profile is well fit by a de Vaucouleurs profile
\citep{0712disc}.  In terms of spectra, the majority  (9--10) of
CLASS lens galaxies have at least one galaxy with an E/S0 spectrum 
\citep{browne93, mckean et al., fassnacht, B1359, 1933disc, B2045_265,
chaeMA, lubin}.  The main exception is B0218+357, which is clearly a
spiral galaxy \citep{browne} and is always excluded from our analysis.
In summary, it is likely that 10-12 of the 13 CLASS lenses contain a
galaxy that would meet the Bernardi et al.\ early-type galaxy selection
criteria (had the SDSS had been able to observe the same redshifts).

The probability that an object is lensed depends on its redshift,
but the redshifts of sources in the CLASS sample are not all known.
We follow \citet{chae et al.} and adopt the following approach:
(i) for lenses, if the source redshift is known it is used,
otherwise $z_S$ is set to the mean value of source redshifts for
the lensed sample, $\langle z_S \rangle = 2$; 
(ii) for unlensed sources, the redshift distribution is modeled
as a Gaussian with $\langle z_S \rangle = 1.27$ and
$\sigma_{z_S}=0.95$, derived from a small subset of the sources
that have measured redshifts \citep{marlow}.
It might be puzzling that the mean redshift for lensed sources
is so much higher than the mean redshift for non-lensed sources;
but Figure~\ref{fig:zdist} shows that the effect is easily explained
by the increase in the optical depth with redshift.

For unlensed sources, we must correct the lensing probability to
account for the resolution limit $\Delta\theta > 0\farcs3$ of the
CLASS survey.  In principle we want to compute the probability of
producing a lens with image separation $\Delta\theta > 0\farcs3$,
although in practice it is more straightforward to compute the
probability of producing any image separation and subtract the
probability of producing an image separation $\Delta\theta <
0\farcs3$, which is what we do.

\begin{figure}[t]
\begin{center}
\epsfig{file=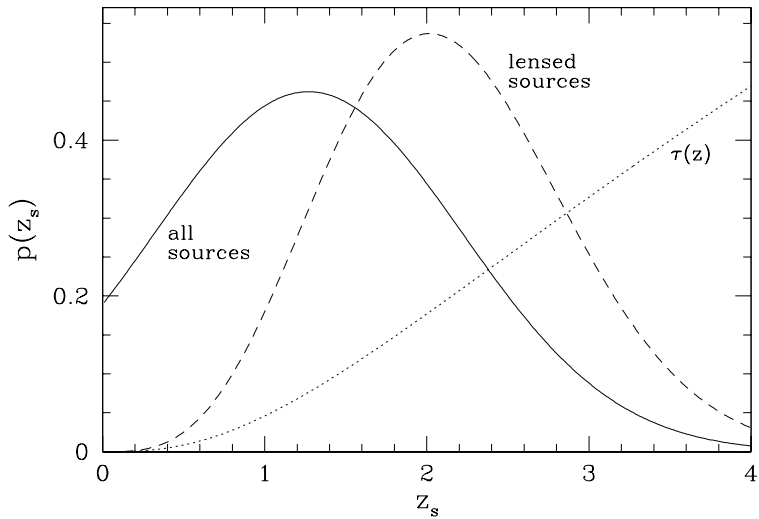,angle=0,width=4.0truein}
\end{center}
\caption{
The solid curve shows the assumed redshift distribution for all
sources in the CLASS survey, $p_{\rm all}(z_s)$.  The dotted curve
shows the optical depth, $\tau(z_s)$, with an arbitrary vertical
scale.  The dashed curve shows the predicted redshift distribution
for lensed sources,
$p_{\rm lensed}(z_s) \propto p_{\rm all}(z_s)\,\tau(z_s)$.  Results
are shown for non-evolving deflector population in a cosmology with
$\Omega_M=0.3$ and $\Omega_\Lambda=0.7$, but are not very sensitive
to these assumptions.
}\label{fig:zdist}
\end{figure}

Figure~\ref{fig:sep} shows the image separation distributions for
the CLASS sample assuming 10 or 12 E/S0 lenses.  Also shown are
the predictions for fiducial models using the SDSS MVF or IVF,
for two different cosmologies: the concordance cosmology,
$\Omega_m = 0.3$ and $\Omega_\Lambda = 0.7$, and our best-fit
cosmology $\Omega_m = 0.9$ and $\Omega_\Lambda = 1.5$ (see \S 4).
The models broadly predict the correct trend in the image separation
distribution, with relatively little sensitively to cosmology.  
Both the MVF and IVF cases predict more sub-arcsecond image
separations than are observed (even using just early-type galaxies),
and hence seem to underestimate the mean separation.  The observed
means are $1\farcs50 \pm 0\farcs14$ and $1\farcs56 \pm 0\farcs15$ 
for the 12 and 10 lens CLASS
samples, respectively (where the errorbars represent an estimate of 
the standard error in the mean, $\sigma/N^{1/2}$).  The MVF model 
predicts a mean separation of $1\farcs35$ for the concordance cosmology, 
or $1\farcs27$ for our best-fit (closed) cosmology.  The IVF model
predicts $1\farcs06$ and $0\farcs99$ for the two cosmologies.

To quantify the apparent disagreement in the separation
distributions, we use the Kolmogorov-Smirnov test to compare the
shapes of the distributions.  We also use a modified version of
the Student $t$-test to compare the means;\footnote{Specifically,
we draw mock samples of 10 or 12 lenses from the model distributions,
compute the means, and determine the fraction of the mock samples
where the mean separation is larger than the observed mean separation.
This is similar to the standard Student $t$-test, except that our
Monte Carlo approach allows us to use the full shape of the model
distribution rather than assuming it to be normal.} this is probably
the most interesting test we can do, because \citet{huterer} show
that it is shifts in the mean image separation that cause the largest
biases in cosmological constraints from lens statistics.  The K-S test
and $t$-test both indicate that the MVF model differs from the data at
no more than 75-95\% confidence.  (The range arises from using the 10
or 12 lens sample and the two cosmologies.)  The IVF model is notably
worse, differing from the data at $>$99\% confidence.  In other
words, the IVF model is sigificantly different from the data, but
the MVF model is only marginally different.

\begin{figure}[t]
\begin{center}
\epsfig{file=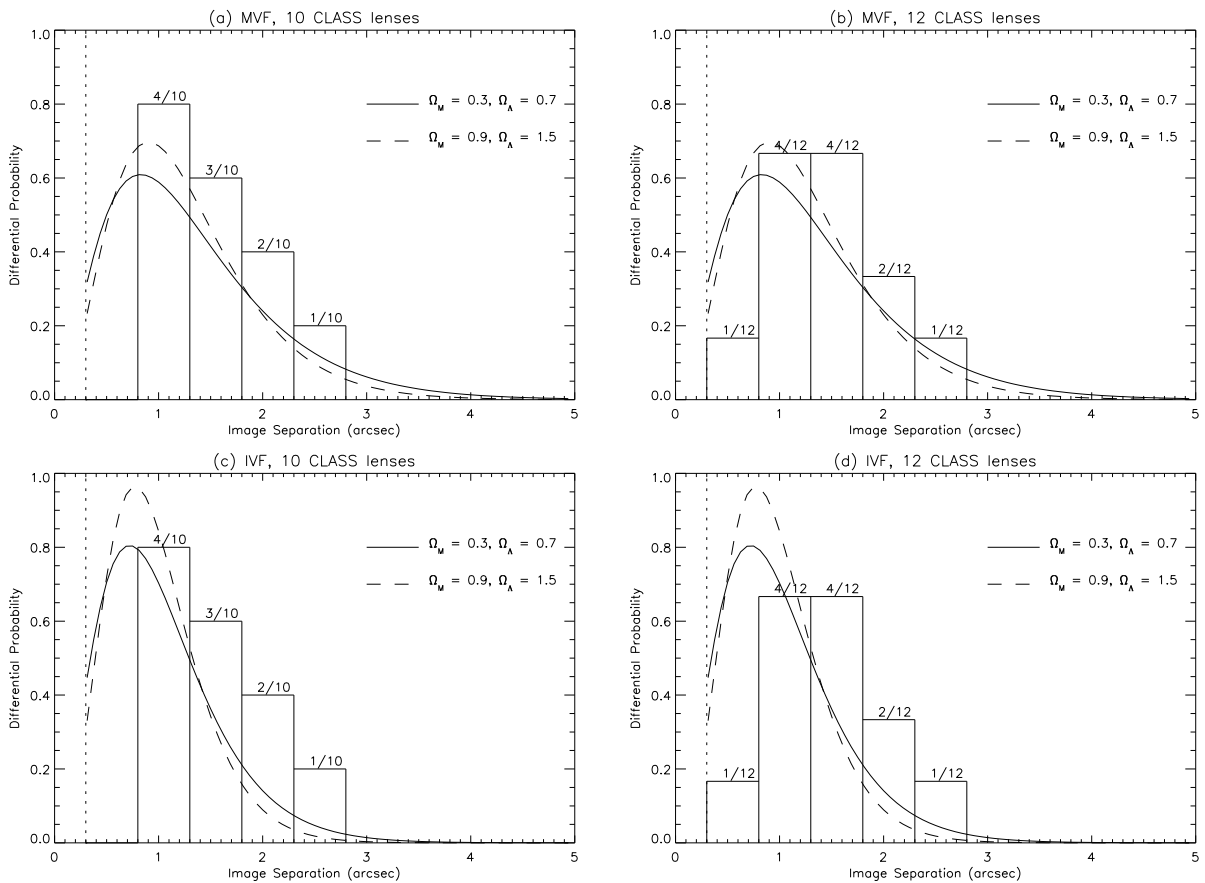,angle=0,width=6.5truein}
\end{center}
\caption{The observed CLASS image separation distribution compared to
predictions based on the SDSS galaxy sample:
(a) MVF and 10 lenses, (b) MVF and 12 lenses,
(c) IVF and 10 lenses, (d) IVF and 12 lenses. 
We show model predictions for two different cosmologies: 
$(\Omega_m,\Omega_\Lambda) = (0.3,0.7)$ and $(0.9,1.5)$.
The dotted line at $\Delta \theta = 0\farcs3$ indicates the CLASS
resolution limit.}
\label{fig:sep}
\end{figure}

\section{Likelihood Analysis of the CLASS Sample}

\subsection{Methods}

In a likelihood analysis, the conditional probability of the data
given a model is the product of the probabilities for the individual
sources.  For an unlensed source, the relevant quantity is the
probability that the source is not lensed, or $(1-\tau)$.  For a
lensed source, the relevant probability depends on the amount of
information that is known about the lens; for example, we can
consider not just the probability that a particular source is
lensed, but rather the probability that it is lensed with a
particular image separation by a galaxy at a particular redshift
(if both $\Delta\theta$ and $z_L$ are known).  Thus, the probability
that enters the likelihood analysis depends on what data are
available:
\be P_{l} = \left\{ \begin{array}{ll}
      \frac{d\tau}{dz_{L}} & \mbox{if $z_{L}$ is known} \\
      \frac{d\tau}{d\Delta\theta} & 
                        \mbox{if $\Delta\theta$ is known} \\
      \frac{d^2\tau}{dz_{L}d\Delta\theta}  
                        & \mbox{if both are known}
      \end{array}.
    \right.  \ee

The conditional probability of the data, $d$, given some model
parameters is then
\be P (d \mid \varepsilon_l,\varepsilon_{c}) = 
\prod^{N_u}_{i=1} (1-\tau^{(i)}) \prod^{N_l}_{j=1} P^{(j)}_l \ .
\label{eq:PL}
\ee
where $N_u$ and $N_l$ are the number of unlensed and lensed sources,
respectively, $\varepsilon_l = (\phi_*,\sigma_*,\alpha,\beta)$ are
the lens model parameters parameters, and
$\varepsilon_c = (\Omega_m,\Omega_\Lambda)$ are the cosmological
parameters.\footnote{Note that the lensing probability does not
depend on the Hubble constant.}
We can incorporate any uncertainties in the lens model parameters
using a prior probability distribution $P(\varepsilon_l)$.  By
Bayes' theorem, the likelihood of the model given the data is then 
\be {\cal L} (\varepsilon_l,\varepsilon_{c} \mid d) = \prod_k 
P(\varepsilon_l^{(k)}) \prod^{N_u}_{i=1} (1-\tau^{(i)}) \prod^{N_l}_{j=1} 
P^{(j)}_l \ . \ee
Because the optical depth is small ($\tau \ll 1$), we can write
\begin{eqnarray} \ln{\cal L} (\varepsilon_l,\varepsilon_{c} \mid d)
&=& \sum_k \ln P(\varepsilon_l^{(k)}) + \sum_{i=1}^{N_u}\ln (1-\tau^{(i)}) + 
\sum_{j=1}^{N_l}\ln P_l^{(j)} \ , \nonumber \\
&\simeq& \sum_k \ln P(\varepsilon_l^{(k)}) - \sum_{i=1}^{N_u}\tau^{(i)}+ 
\sum_{j=1}^{N_l}\ln P_l^{(j)} \ , \nonumber \\
&\simeq& \sum_k \ln P(\varepsilon_l^{(k)}) - \int N(z_S)\,\tau(z_S)\,dz_s + 
\sum_{j=1}^{N_l}\ln P_l^{(j)} \ , 
\label{eq:logL}
\end{eqnarray} 
where $N(z_S)$ is the redshift distribution of CLASS sources 
(see \S 3.2), normalized to the number of unlensed sources in the 
statistical sample.  

In principle, a likelihood analysis of lens statistics can be used
to probe either the lens galaxy population \citep[e.g.,][]{davis}
or cosmology.  We focus on the latter and marginalize over lens
model parameters as appropriate.  When using the measured velocity
function, we find that uncertainties in the MVF parameters have
negligible effect on cosmological conclusions (see \S 4.3).  When
using the inferred velocity function, the most important uncertainty
is in $\sigma_*$, partly because the optical depth is so sensitive
to $\sigma_*$ (see eqn.~\ref{eq:tauMVF}), and partly because the
scatter in the Faber-Jackson relation effectively leads to a large
uncertainty in $\sigma_*$.  We combine the inverse Faber-Jackson
relation and its scatter, eqns.~\ref{eq:ifj} and \ref{eq:ifjscatt},
with $M^*_r$ from the Schechter luminosity function, to obtain a
Gaussian prior on $\sigma_*$.  We then marginalize over $\sigma_*$:
\be {\cal L} (\varepsilon_{c} \mid d) = \int
  {\cal L} (\varepsilon_l, \sigma_* \mid d)\ d\sigma_*\ .
\ee
In this analysis, we keep  the power-law index $\gamma$ of the
Faber-Jackson relation fixed at the best-fit value, $\gamma = 4.4$.
We also assume the luminosity function parameters in 
eqn.~\ref{eq:LF} are well determined and fix them at their best-fit
values.  These assumptions are justified because the uncertainties
in the luminosity function parameters are small compared to the
scatter in the Faber-Jackson relation.

\subsection{Cosmological constraints: no-evolution model}

\begin{figure}[t]
\begin{center}
\epsfig{file=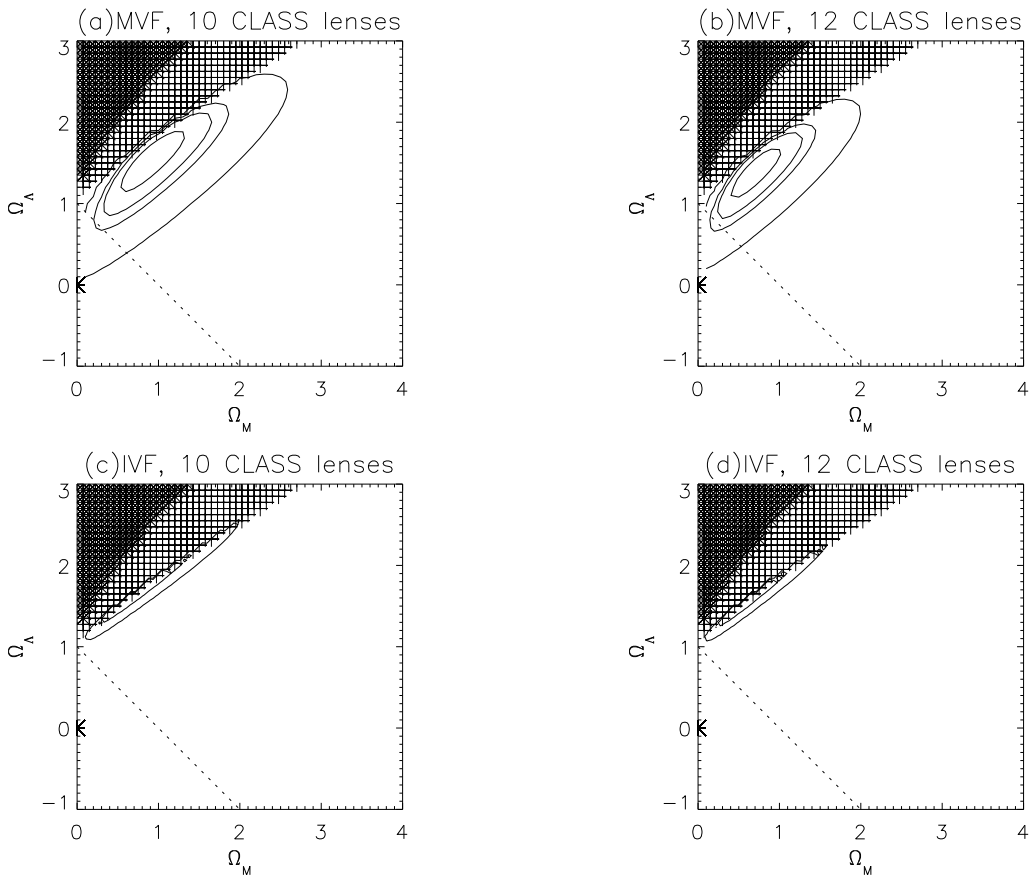,angle=0,width=6.5truein}
\end{center}
\caption{Likelihood contours for the SDSS inferred velocity function
(IVF) and measured velocity function (MVF) for 10 and 12 CLASS lenses.  
Contours are drawn at the 68, 90, 95, and 99$\%$ confidence levels for
the MVF model; but only the 95 and 99\% contours for the IVF model,
because the other contours run into the shaded region where the
cosmology is either unphysical (has imaginary comoving distances, dark 
shaded region) or has no Big Bang (a bounce at $z<6$, light shaded region).  
The dotted line marks spatially flat cosmologies.
} 
\label{fig:Lplot}
\end{figure}

\begin{figure}[t]
\epsfig{file=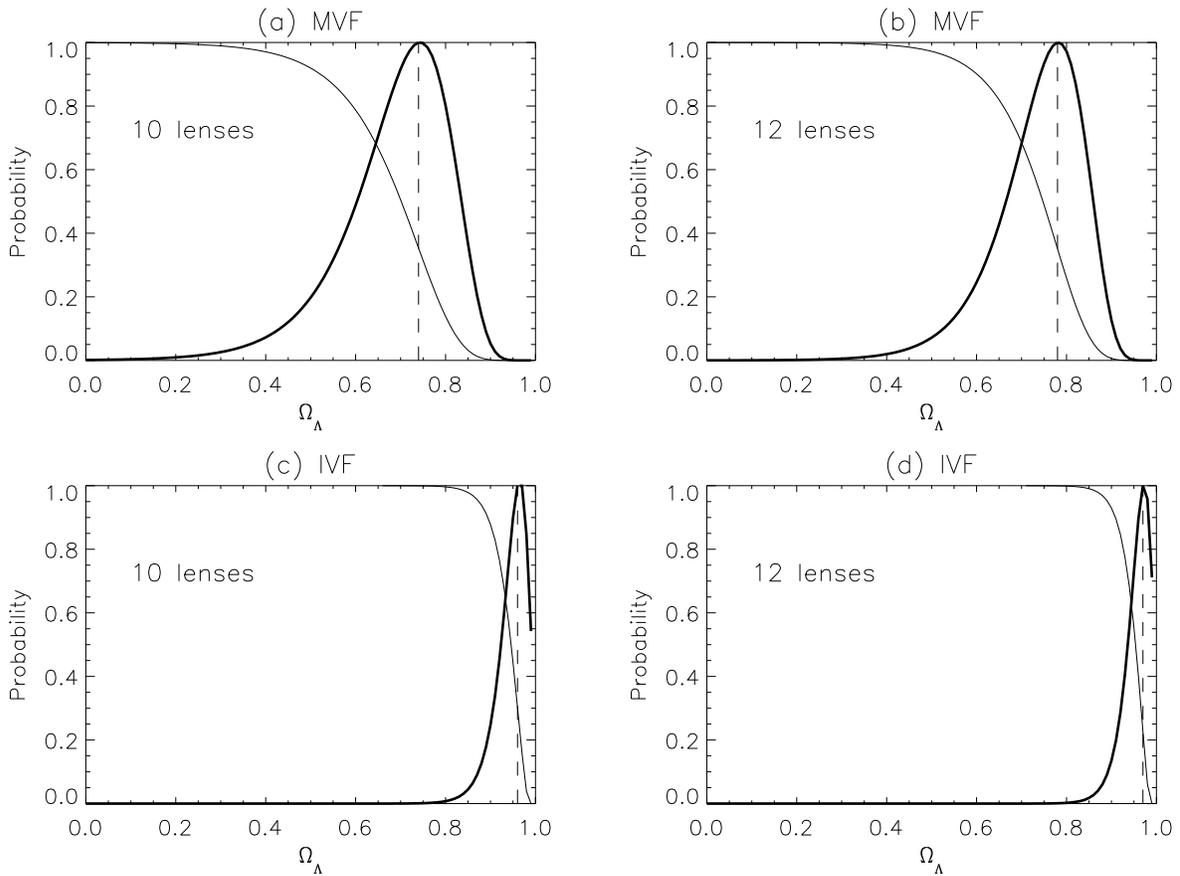,angle=0,width=6.5truein}
\caption{Slices of relative probability along the line of spatially
flat cosmologies, for the four models in Fig.~\ref{fig:Lplot}.  The
thick curves show the differential probabilities, $dP/d\Omega_\Lambda$,
while the thin curves show the cumulative probabilities, 
$P(>\!\Omega_\Lambda)$.  The dashed lines mark the maximum likelihood
values.}
\label{fig:flat}
\end{figure}

We first follow many of the previous analyses of lens statistics
and assume that the velocity function does not evolve.
Figure~\ref{fig:Lplot} shows likelihood contours in the plane of
$(\Omega_m,\Omega_\Lambda)$ using the CLASS sample with either
10 or 12 early-type lenses, and using either the SDSS MVF or IVF
lens model parameters.  For the 12 lens sample the most likely values
are $\Omega_m = 0.9$ and $\Omega_\Lambda = 1.5$ for 12 CLASS lenses,
while for the 10 lens sample the most likely values lie outside the
range of physical cosmologies.  Figure~\ref{fig:flat} shows the
relative likelihood versus $\Omega_\Lambda$ along the slice through
the $(\Omega_m,\Omega_\Lambda)$ plane corresponding to a spatially
flat cosmology ($\Omega_m + \Omega_\Lambda = 1$).  Table~\ref{table2}
gives quantitative constraints on $\Omega_\Lambda$ for flat
cosmologies.

We mentioned in \S 3.1 that neglecting the scatter in the
Faber-Jackson relation causes the IVF model to underestimate the
abundance of massive early-type galaxies, and hence underestimate
the lensing optical depth.  This causes a significant bias toward
{\it higher\/} values of $\Omega_\Lambda$.  The shift between the
IVF and MVF models is $\Delta\Omega_\Lambda \simeq 0.2$ for flat
cosmologies, which pushes $\Omega_\Lambda$ disturbingly close to
unity.  More generally, the IVF model requires a cosmology with a
very large dark energy component that borders on being unphysical.
The scatter in the Faber-Jackson relation is clearly important for
lens statistics.

\begin{table}[t]
\begin{center}
\begin{tabular}{ l |c c c c| c c c c }
\hline
Model & \multicolumn{4}{c}{10 CLASS early-type lenses} & 
        \multicolumn{4}{c}{12 CLASS early-type lenses} \\
\cline{2-9}
  & MLE   & 68\%   & 95\%   & UL  
  & MLE   & 68\%   & 95\%   & UL   \\
\hline
IVF  & $0.96$ & $^{+0.03}_{-0.03}$   & $^{\rm\ NA}_{-0.06}$ & NA
     & $0.97$ & $^{\rm\ NA}_{-0.03}$ & $^{\rm\ NA}_{-0.06}$ & NA \\
&&&&&&&\\
MVF  & $0.74$ & $^{+0.09}_{-0.11}$   & $^{+0.14}_{-0.28}$   & $0.84$  
     & $0.78$ & $^{+0.07}_{-0.10}$   & $^{+0.12}_{-0.23}$   & $0.86$ \\
&&&&&&&\\
eMVF & $0.72$ & $^{+0.13}_{-0.18}$   & $^{+0.20}_{-0.46}$   & $0.86$    
     & $0.78$ & $^{+0.10}_{-0.16}$   & $^{+0.16}_{-0.38}$   & $0.89$ \\
\hline
\end{tabular}
\end{center}
\caption{Constraints on $\Omega_\Lambda$ for spatially flat
cosmologies, using models based on the IVF, the MVF (neglecting
evolution), and the MVF including the effects of evolution (`eMVF').
We quote the maximum likelihood estimate (`MLE'), the 68\% and 95\%
confidence limits, and the 95\% confidence upper limit (`UL').  We
give results for cases with 10 or 12 CLASS E/S0 lenses.}
\label{table2}
\end{table}

Note the curious result that the IVF model appears to yield tighter
cosmological constraints than the MVF model, even though we have
included uncertainty in $\sigma_*$ in the IVF analysis.  The
difference can be explained by the dependence of the comoving
volume element on the cosmological parameters.  Poisson errors
in the lens sample can be thought of as giving some particular
uncertainty $\sigma_V$ in the cosmological volume.  The inferred
uncertainty in $\Omega_\Lambda$ is, conceptually,
\be \sigma_{\Omega_\Lambda} = \frac{\sigma_V}{dV/d\Omega_\Lambda}\ . \ee
The derivative $dV/d\Omega_\Lambda$ increases rapidly as
$\Omega_\Lambda$ increases, leading to a {\it decreasing\/}
uncertainty $\sigma_{\Omega_\Lambda}$.  Because the IVF has a
larger best-fit value of $\Omega_\Lambda$ than the MVF, it has
a smaller inferred uncertainty.

It is difficult to compare our results directly with those of 
\citet{chae} and \citet{chae et al.}, since they find that
uncertainties in the late-type galaxy population lead to
considerable uncertainties in the cosmological constraints.
(As mentioned in \S 3.1, this is a large part of our rationale
for excluding late-type lenses from our analysis.)  Depending on
priors placed on the late-type population, \citet{chae} finds
best-fit values of $\Omega_\Lambda$ for a flat Universe between
0.60 and 0.69.  These values are $\sim$0.1 lower than ours
because the SSRS2 IVF produces a higher optical depth than the
SDSS MVF (see Fig.~\ref{fig:LE}).

The constraints in the $(\Omega_m,\Omega_\Lambda)$ plane from
the MVF model are qualitatively similar in shape and orientation
to those derived from the redshift-magnitude relation in Type Ia
supernovae \citep[e.g.,][]{tonry}.  The reason is that both phenomena
measure cosmological distances at moderate redshifts $z \sim 1$.  One
of the key results from Figure~\ref{fig:Lplot} is that lensing requires
$\Omega_\Lambda>0$ at more than 99\% confidence, even without assuming
a flat universe.  This is important confirmation of the evidence from
supernovae that there is a nonzero dark energy component in the
universe.

\subsection{Effects of statistical uncertainties in the MVF parameters}

In the previous section we assumed that the MVF parameters were
known precisely.  To consider how statistical uncertainties in the parameters
affect the cosmological constraints, we adopt a Monte Carlo
approach that automatically includes important covariances between
the parameters.  Specifically, we created 1000 mock catalogs each
containing 30,000 velocity dispersions drawn from the best Schechter
function fit to the SDSS MVF.  We then refit each catalog to produce
1000 sets of lens model parameters that represent the scatter and
covariance associated with having a finite number of galaxies.
This is identical to the procedure used by \citet{sheth et al.}
to estimate the uncertainties in the MVF parameters.

We then repeated the likelihood analysis of the CLASS sample using
the 1000 sets of mock lens parameters.  Figure~\ref{fig:spread}
shows the resulting maximum likelihood estimates of $\Omega_m$ and
$\Omega_\Lambda$.  The statistical uncertainties in the MVF parameters clearly
have a negligible effect on the cosmological constraints, producing
a scatter of just $\sim$0.006 in $\Omega_m$ and $\sim$0.010 in 
$\Omega_\Lambda$.  Other systematic uncertainties are
discussed in \S4.5.

\begin{figure}[t]
\begin{center}
\epsfig{file=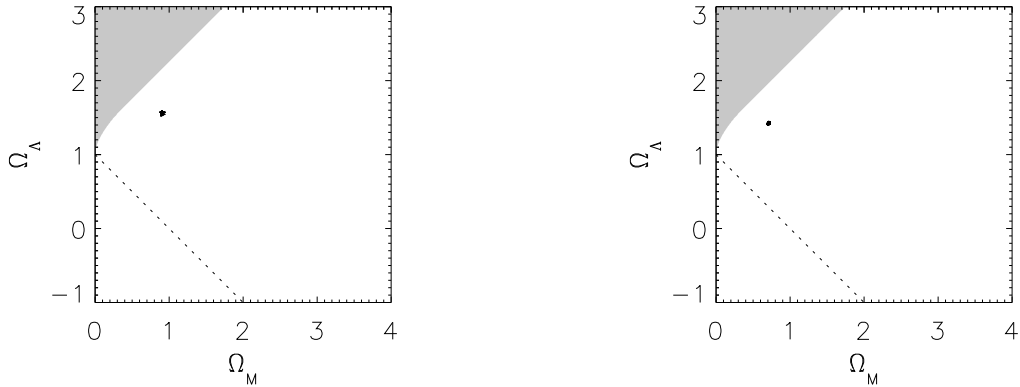,angle=0,width=6.5truein}
\end{center}
\caption{The scatter in the maximum likelihood estimates of
$\Omega_m$ and $\Omega_\Lambda$ due to uncertainties in the MVF
parameters, based on 1000 Monte Carlo realizations of the SDSS MVF.
We show results for 10 and 12 CLASS E/S0 lenses (left and right
panels, respectively).}
\label{fig:spread}
\end{figure}

\subsection{Effects of evolution}

We now consider how evolution in the velocity function can affect
the cosmological constraints we derive.  We do this by using the theoretical 
evolution model described in Section 2.3 together with the SDSS MVF.  
(In our model, the VF evolves substantially less than advocated by 
\citet{chae-evol}.  However, our model for evolution is more in 
line with the findings of \citet{ofek}.)  Figure~\ref{fig:evol} 
shows the probability versus $\Omega_\Lambda$
for spatially flat cosmologies.  The maximum likelihood estimate
and 1$\sigma$ uncertainties are $\Omega_\Lambda = 0.72^{+0.13}_{-0.18}$
for 10 CLASS E/S0 lenses, or $0.78^{+0.10}_{-0.16}$ for 12 E/S0
lenses (see Table~\ref{table2}).
The image separation distribution for our evolving model is not 
significantly different than our non-evolving model for flat 
cosmologies.

\begin{figure}[t]
\epsfig{file=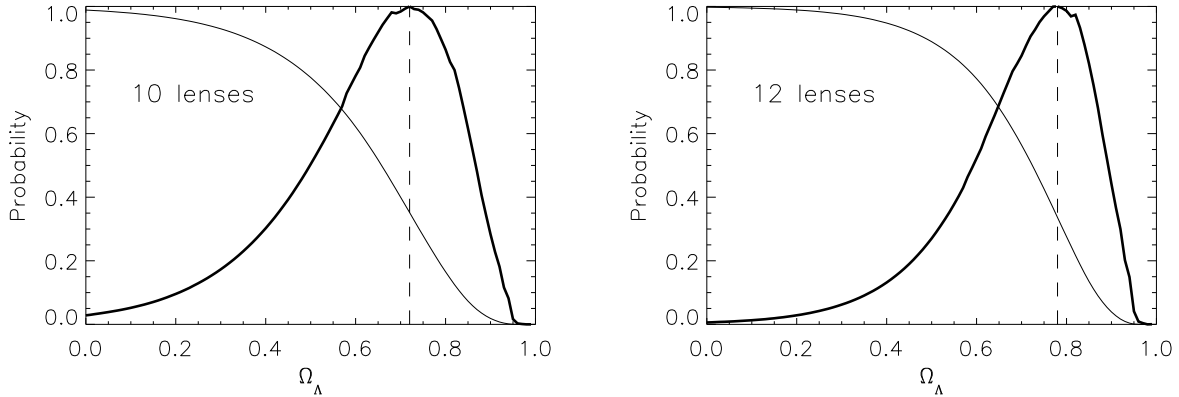,angle=0,width=6.5truein}
\caption{Relative probability versus $\Omega_\Lambda$ for spatially
flat cosmologies, using the SDSS MVF and our evolution model based
on extended Press-Schechter theory.}
\label{fig:evol}
\end{figure} 

Surprisingly, evolution appears to broaden the uncertainties on
$\Omega_\Lambda$ without shifting the maximum likelihood value.
The increase in the uncertainties is fairly straightforward to
understand.  The evolution model predicts that $\phi(\sigma)$
increases between $z=0$ and $z=1$ (except for rare, very massive
galaxies; see Fig.~\ref{fig:ratio}), which would increase the
optical depth.  But the effect weakens as $\Omega_\Lambda$ increases,
which partially offsets the increase in the cosmological volume
and causes $\tau(\Omega_\Lambda)$ to be less steep for the
evolution model than for the no-evolution model.  The Poisson
errors in the lens sample (or, equivalently, in the measured value
of $\tau$) therefore translate into larger uncertainties in
$\Omega_\Lambda$.  Our results confirm the suggestion by
\citet{keeton02a} that cosmology dependence in the evolution rate
can weaken the cosmological conclusions drawn from lens statistics.

Understanding why evolution produces no shift in the maximum
likelihood values is more subtle.  Because the velocity function
is predicted to rise from $z=0$ to $z \sim 1$ (over the relevant
range of $\sigma$; see Fig.~\ref{fig:ratio}), we might naively
expect that evolution would increase the optical depth and push
us to lower values of $\Omega_\Lambda$.  However, there are
actually competing effects in the likelihood.  Consider the
expression for the log likelihood in eqn.~\ref{eq:logL}.  The
maximum likelihood corresponds to the point where the derivative
with respect to $\Omega_\Lambda$ vanishes --- or where the
derivatives of the first and second terms in eqn.~\ref{eq:logL}
are equal.  Figure~\ref{fig:slopes} shows these two derivatives
as a function of $\Omega_\Lambda$, for both non-evolving and
evolving MVF models.  As just mentioned, evolution flattens the
dependence of the optical depth on $\Omega_\Lambda$, lowering the
derivatives.  But it affects the two terms differently, because
the lens term is a sum of $\log\tau$ while the non-lens term is
a sum of $\tau$ itself.  The flattening effect fortuitously
cancels near $\Omega_\Lambda \simeq 0.78$, so there is no shift
in the location of the maximum likelihood.  We emphasize that
the almost perfect cancellation near the concordance cosmology
is a coincidence; if the best-fit value of $\Omega_\Lambda$ were
something different, then we would see evolution produce a shift
in the location of maximum likelihood.  But as it stands,
evolution does not appear to have a strong effect on our
cosmological constraints.  This result
is consistent with the conclusions of \citet{rix94} and
\citet{mao and kochanek}, who used simple evolution models to
argue that mergers do not significantly affect lens statistics if the
progenitor and product galaxies all lie on the fundamental plane of
early-type galaxies (i.e., if the kinematics features are conserved).

\begin{figure}[t]
\begin{center}
\epsfig{file=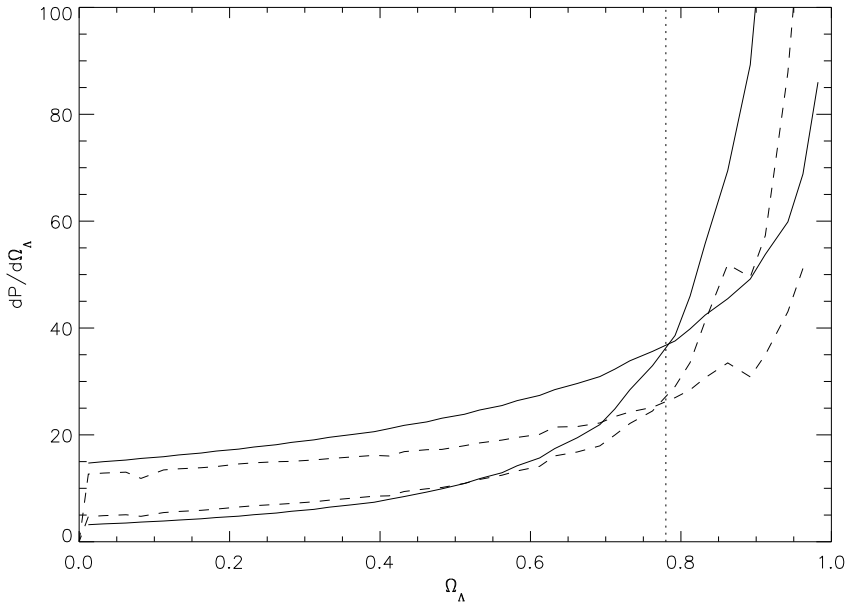,angle=0,width=4.7truein}
\end{center}
\caption{Derivatives of the two $\Omega_\Lambda$-dependent terms
in the log likelihood, eqn.~\ref{eq:logL}, for our MVF (solid lines)
and evolving MVF (dashed lines) models.  In each case the lower and
upper curves represent the terms for non-lensed and lensed sources,
respectively; the likelihood is maximized at the point where the
curves cross.  Jaggedness in the curves is due to numerical noise.
We show results for 12 CLASS E/S0 lenses.  The dotted line marks
$\Omega_\Lambda = 0.78$.}
\label{fig:slopes}
\end{figure}

\subsection{Other systematic effects}

We believe that by improving the model of the deflector population
and considering possible redshift evolution, we have dealt with
the major systematic uncertainties in lens statistics constraints
on flat cosmologies.  There are, however, some additional effects
that should be discussed.  Further data and/or analysis will be
required to account for them fully, but we can identify the
direction and estimate the amplitude of the effects.

First, recall that we have assumed spherical deflectors.
\citet{huterer} have recently studied how lens statistics are
affected by ellipticity in lens galaxies and external tidal shear
from neighboring objects.  They find that reasonable distributions
of ellipticity and shear have surprisingly little effect on both
the total optical depth and the mean image separation.  Ellipticity
and shear do broaden the image separation distribution slightly,
but that effect is not important for cosmological constraints.
In particular, \citet{huterer} quantify the biases in cosmological
constraints due to neglecting ellipticity and shear.  They find
errors of $\Delta\Omega_M = 0.00\pm0.01$ and
$\Delta\Omega_\Lambda = -0.02\pm0.01$ (where the errorbars indicate
statistical uncertainties from the Monte Carlo calculation method).
In other words, ellipticity and shear either do not affect the
parameters at all or affect them at a level that is unimportant.

Second, recent work has suggested that neglecting lens galaxy
environments can bias lens statistics.  Satellite galaxies
\citep{cohn} and groups or clusters around lens galaxies
\citep{kz04} can increase lens image separations and cross
sections.  Conversely, neglecting their effects (as we and
nearly all other authors have done) can cause underestimates of
the image separations and cross sections, and hence overestimates
of $\Omega_\Lambda$.  Poor knowledge of the distribution of lens
galaxy environments prevents a detailed calculation of the effect,
but \citet{kz04} estimate that the shift in $\Omega_\Lambda$ for
flat cosmologies is certainly less than 0.14 and more likely to
be at the level of $\sim$0.05.  Surveys to characterize lens
environments are now underway, and they will make it possible
to account for this effect in future lens statistics calculations.

Third, the Appendix suggests
the possibility that as much as 30\% of the early-type galaxy population
was excluded by the Bernardi et al. (2004) sample selection, although the
fraction is probably much smaller.  If we assume the maximum omitted
fraction, then our estimates of $\Omega_\Lambda$ for a flat cosmology
would drop by $\sim$0.05 for the IVF model and $\sim$0.15 for the MVF
model.

Finally, the most significant limitation of the CLASS sample is
poor knowledge of the source redshift distribution.  \citet{chae}
estimates that this leads to an uncertainty (but not necessarily
a bias) of $\sim$0.11 in $\Omega_\Lambda$.

\section{Conclusions}

We have derived new constraints on the cosmological parameters 
using the statistics of strong gravitational lenses.  We have
modified lens statistics calculations in two important ways.
First, we point out that neglecting scatter in the Faber-Jackson
relation biased the results of previous analyses of lens statistics
\citep[also see][]{kochanek94}.  Working with a direct measurement
of the velocity dispersion distribution function removes these
biases.  Second, we use a theoretical model for the redshift
evolution of the velocity function to study how evolution affects
lens statistics.  These modifications allow us to obtain more
robust cosmological constraints.

We find good agreement between lens statistics and the current
concordance cosmology (at 1$\sigma$) and with the recent results
from Type Ia supernovae \citep[e.g.,][]{tonry}.  Our maximum
likelihood flat cosmology for the (non-evolving) MVF model has
$\Omega_\Lambda=0.74^{+0.09}_{-0.11}$ if 10 of the 13 CLASS
lenses are produced by early-type galaxies, or
$\Omega_\Lambda=0.78^{+0.07}_{-0.10}$ if there are 12 CLASS
early-type lenses.  Neglecting the scatter in the Faber-Jackson
relation (using the IVF rather than the MVF) would bias the
results toward higher values of $\Omega_\Lambda$, with a shift
$\Delta\Omega_\Lambda \simeq 0.2$ that is twice as large as the
statistical errors.  If there is evolution in the velocity
function that can be modeled with extended Press-Schechter theory,
it has surprisingly little effect on the maximum likelihood values
of $\Omega_\Lambda$ but it does increase the uncertainties by
$\sim$50\%.  The Appendix suggests the \citet{bernardi2} sample 
might be missing up to 30\% of early-type galaxies, but this omitted 
fraction is likely much smaller.  If the full 30\% are being omitted, then 
our estimates of $\Omega_\Lambda$ for a flat cosmology would drop 
by $\sim$0.05 for the IVF model and $\sim$0.15 for the MVF model.

While it is gratifying to see that lens statistics now agree with
what are considered to be strong cosmological constraints from
supernovae and the cosmic microwave background, one may wonder
whether the lensing results are actually interesting.  We believe
that they are, for several reasons.  Perhaps the most essential
question in cosmology today is whether there is a dark energy
component.  To date the only single dataset able to address that
question has been the supernovae.  (The CMB constrains the total
density $\Omega_\Lambda+\Omega_m$, while clusters constrain
$\Omega_m$.)  Perhaps the most significant result from lens
statistics is strong evidence for $\Omega_\Lambda>0$, absent
any other cosmological assumptions (see Fig.~\ref{fig:Lplot}).
With the underlying physics of Type Ia supernovae not understood,
the confirmation from lensing is significant.  
Alternatively, if the cosmology is known and accepted from other
methods, then lensing will provide perhaps the cleanest probe of
dynamical evolution in the early-type galaxy population to test
the paradigm of hierarchical structure formation that forms the
other main pillar of our cosmological paradigm.

\acknowledgements

We thank
M.~Bernardi for providing data in advance of publication,
A. Kravtsov for supplying the simulation results used in 
Figure \ref{fig:ratio}, K.~Chae for communications about the 
CLASS/SSRS2 analysis, J.~Newman for discussions about the 
evolution model, and K.~Wieand for conversations about statistical 
tests.  We thank D.~Rusin, C.~Kochanek, and A.~Zabludoff for
interesting general discussions.
We thank the anonymous referee for detailed and helpful
comments on the manuscript.  We also thank D. Huterer for a 
careful reading of the manuscript and L. Krauss for helpful 
comments.
We acknowledge support from
the NSF Center for Cosmological Physics at Chicago, from
the DOE,
from NASA Hubble Fellowship grant HST-HF-01141.01-A at Chicago,
and from NASA grant NAG5-10842 at Fermilab.


\pagebreak

\appendix

\section{Selection of the sample of early-type galaxies}

The selection procedures outlined by \citet{bernardi} define the 
sample we use in our analysis.  This Appendix studies the roles
played by each step in the selection process.  We find that the
sample could plausibly underestimate the true abundances of massive
early-type galaxies of interest for lensing, but not by more than
$\sim$30\%.  We then argue that the missed fraction is likely to be
considerably smaller, based on recent measurements \citep{cross} of
the early-type galaxy luminosity function at the redshifts
$z \sim 0.5$--1 relevant for lensing.

We begin with an observation that SDSS data have made quite clear:
to a rather surprising approximation, the galaxy distribution is
bimodal (e.g., \citet{blanton3}).  \citet{baldry} describe
how the $u-r$ vs $r$ color-magnitude diagram can be used to construct
an optimal division between what are essentially red and blue
populations.  Since it is widely accepted that giant early-type
galaxies are red, they almost certainly belong to the red population.
Of course, the red population may also have a substantial number of
edge-on disks, so the bimodality in color almost certainly does not
translate simply into a bimodality in morphology.  Nevertheless, we
will use this red population as a basis against which to compare the
\citet{bernardi} selection process.  

We do this by selecting objects from the SDSS main galaxy sample, 
restricted to the range $13.5\le r_{\rm Pet}\le 17.77$ and $z\le 0.08$, 
following \citet{baldry}.  
This gives 71,517 objects.  The crosses in Figure~\ref{fig:phiselect} show our estimate of the 
luminosity function of this sample; it is well-described by the solid 
line, which shows the estimate published by \citet{baldry}  
Selecting with the color cuts in \citet{baldry} (where all magnitudes
are Petrosian),
\begin{displaymath}
 M_u-M_r>2.06 - 0.244\tanh[(M_r+20.07)/1.09],
\end{displaymath} 
yields a red subsamble containing 29908 objects. The crosses show our estimate of the red galaxy luminosity function, 
and the dashed line shows $\phi_{red}$ published by \citet{baldry}  
Once again the agreement is reassuring, suggesting that our estimator 
of the luminosity function is accurate.  

\begin{figure}[t]
\begin{center}
\epsfig{file=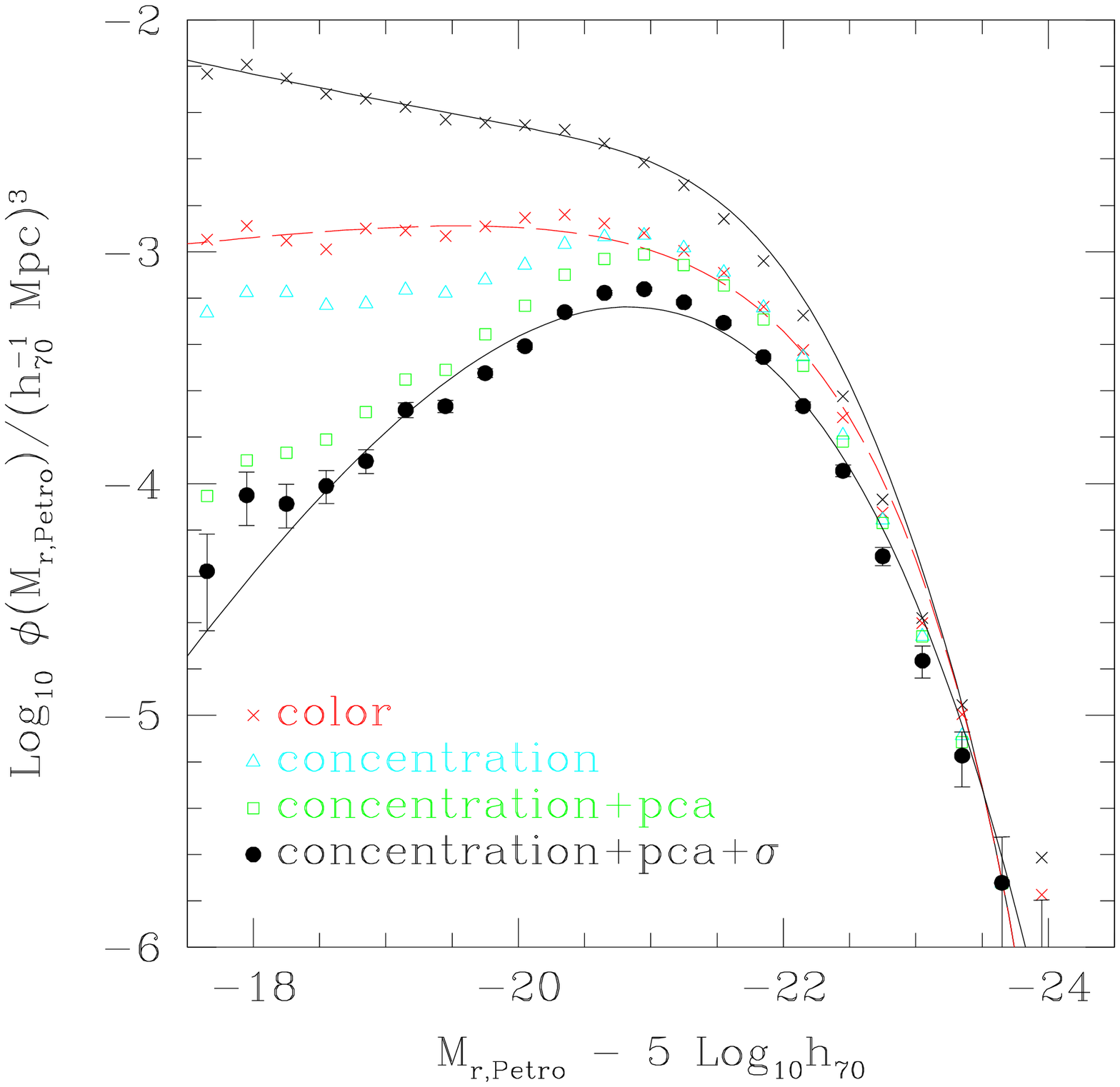,angle=0,width=2.9truein}
\epsfig{file=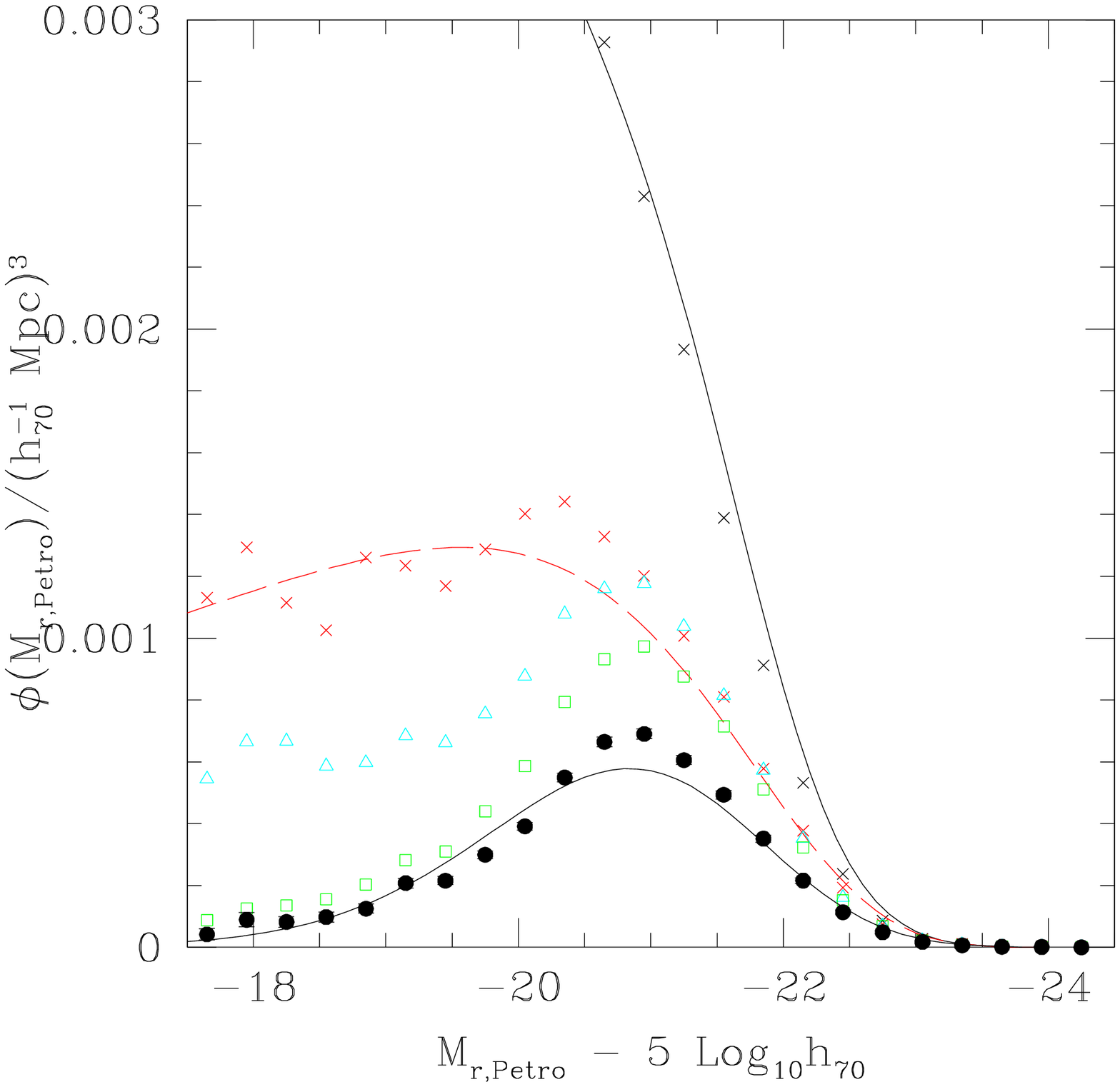,angle=0,width=2.9truein}
\end{center}
\caption{Luminosity functions associated with various subsamples 
 drawn from the SDSS main galaxy sample at $z\le 0.08$ and 
 $13.5\le r_{\rm Pet}\le 17.77$ .  Top set of crosses shows the 
 luminosity function of the full sample, and the lower set is for the 
 red subsample defined by \citet{baldry}; curves show their 
 published fits to $\phi_{tot}$ and $\phi_{red}$.  
 Triangles, squares and circles show subsamples defined by applying 
 successive cuts in concentration, spectral type, and velocity 
 dispersion.  Bottom curve shows equation~(\ref{eq:LF}), shifted 
 faintwards by 0.1~mags to account for the fact that Petrosian magnitudes 
 used here only account for about 90\% of the light in a de Vaucouleurs 
 profile.  
 In all cases, the luminosity function has been estimated by assuming 
 that the sample is purely magnitude limited.}
\label{fig:phiselect}
\end{figure}

Next, we compare this sample of red galaxies with the sample selected 
following \citet{bernardi}.  Because one of their goals was to 
study the color magnitude relation of early-type galaxies, Bernardi et
al.\ did not select on color.  Instead, they used a combination of
photometric and spectroscopic cuts to define their sample.   To mimic
their selection, we return to the full galaxy sample, apply the same
redshift and apparent magnitude cuts, and then select all objects for
which the concentration index of the light profile in the $i_{\rm Pet}$
band photometry is greater than $2.75 - 5z/6$, where $z$ is the redshift.  
Use of the concentration as an indicator of galaxy type is motivated by 
\citet{strateva}, and this concentration cut is one of the cuts 
made by Bernardi et al.  The redshift dependence is included to account 
for the fact that the concentration index is not seeing-corrected.  
The triangles show the luminosity function associated with the 23,857 galaxies which satisfy this cut:  notice that it tracks $\phi_{red}$ well 
at the most luminous end, but that it is lower by about 0.3 dex at lower 
luminosities ($M_{r,{\rm Pet}}>-21$).  Direct inspection of the images of 
a random sample of the red objects which do not satisfy the concentration 
cut shows that they are predominantly edge-on disks.  Thus, for the 
purposes of selecting an early-type galaxy sample, the cut on 
concentration is more efficient and accurate than selecting on color.  

We then applied the \citet{bernardi} cut on spectral-type, obtained from 
a PCA analysis of the spectrum:  the specific requirement is that the 
spectroscopic pipeline parameter {\tt eclass}$<0$.   This is essentially 
a cut on the shape of the continuum, and removes objects whose spectra 
indicate recent star-formation.  Open squares show the luminosity 
function of the 17,977 objects that remain; the cut on spectral type
removes many more of the lowest luminosity objects, but makes little
difference at the luminous end which is most relevant for the present
study.  

Finally, filled circles show the luminosity function for the subset of 
12,490 objects which satisfied both the concentration and spectral type cuts, and for which the spectroscopic pipeline reports a measured velocity dispersion.  Velocity dispersions are only measured if the signal-to-noise ratio of pixels in particular wavelength intervals of the spectra is 
sufficiently high, so this final cut does not have any underlying physical 
motivation --- it is made purely so that measured velocity dispersions are 
reliable.  The filled circles fall about 0.2 dex below the open squares 
at small luminosities, but they are quite similar to the squares at 
the highest luminosities.  This suggests that, by only including objects 
for which reliable velocity dispersion measurements are available, 
\citet{bernardi} may have removed bona-fide low and moderate 
luminosity early-type galaxies from the sample, but it has not 
significantly reduced the inferred abundance of the most luminous 
objects.  Thus, while the number density of early-type galaxies may be 
larger than they quote, but it is unlikely to be more than 
$10^{0.2}\sim 50\%$ higher at $L<L_*$, and it is likely to be unchanged 
at $L>6L_*$.  

If the \citet{bernardi2} sample is missing some bona-fide early-types, 
then one way to quantify this is to estimate the fraction of the 
total luminosity density contributed by early-types.  
The luminosity density in the \citet{baldry} red sample is 42\% of the 
total (consistent with the red galaxy sample of \citet{hogg}, 
but this is almost certainly an overestimate of the contribution from 
early-type galaxies to the luminosity density.  
The cut on concentration rather than color leaves 39\% of the total 
luminosity density, including the cut on spectral type reduces this 
to 31\%, and requiring that reliable $\sigma$ measurements are 
available reduces this to 22\%.  These numbers are actually slight 
underestimates of the early-type contribution, because they were 
computed from the Petrosian luminosity, whereas the luminosity of an 
early-type galaxy is actually better represented by the de Vaucouleurs 
luminosity.  As \citet{blanton1} demonstrate, the Petrosian 
luminosity underestimates the true value by about 10\%.  
If $j_{\rm Petro} = e/(e+l)$, 
 then correcting for this difference means that 
 $j_{\rm deV} = 1.1e/(1.1e + l) = 1/(1 + l/1.1e) 
              = j_{\rm Petro} 1.1/(1 + 0.1j_{\rm Petro})$.  
Hence, the early-type fractions associated with these cuts, 
0.39, 0.31 and 0.22, become 0.42, 0.34, and 0.24.  
Since the final cut does not change the shape of the luminosity 
distribution drastically (the squares and circles trace curves 
of similar shape), it may be that the final cut leaves a sample which 
underestimates the true number density of ellipticals by about 30\%\
(i.e., $1-0.24/0.34$).  Therefore, it may be that the distribution of
velocity dispersions $\phi(\sigma)$ in the Bernardi et al.\ sample
accurately represents the true shape of the MVF, but underestimates
the number density by a similar factor.  

There is one important caveat, however:  the analysis above is
restricted to low redshifts.  More relevant for our analysis is
the distribution of velocity dispersions at typical lens redshifts,
$z \sim 0.5$--1.  In the IVF model, we use the $\phi(L)$ distribution
measured at low redshifts and assume that the only change with
redshift is pure luminosity evolution.  We can test this assumption
by comparing the luminosity function we use with that recently measured
from 48~arcmin$^2$ of HST-ACS fields at $z\sim 0.75$ \citep{cross}.
Figure~\ref{fig:phiacs} illustrates that the number densities from 
the higher-redshift sample are slightly lower than those from the 
SDSS `concentration+pca+$\sigma$' sample, and substantially lower than 
those from the SDSS sample in which the cut on $\sigma$ has been 
omitted.  Therefore, using all three cuts to define the early-type
sample is probably a better approximation than using only the
`concentration+pca' cuts.

\begin{figure}[t]
\begin{center}
\epsfig{file=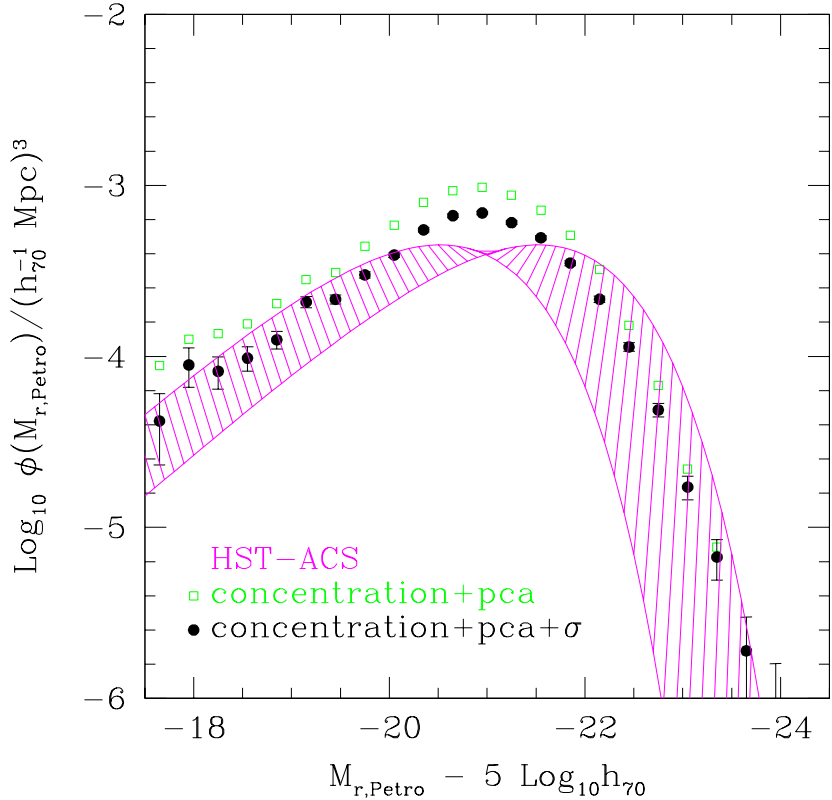,angle=0,width=4.7truein}
\end{center}
\caption{Early-type galaxy luminosity functions from the SDSS 
 sample (squares and circles show subsamples defined by applying 
 successive cuts in concentration, spectral type, and velocity 
 dispersion, as in previous figure), and from the HST-ACS survey 
 (hashed region shows the luminosity function of objects with 
 $(U-V)>1.38$ and $0.5\le z\le 0.75$, shifted to $z=0$ assuming 
 $M_r = M_B-1.32-1.76z_{ACS}$ with $z_{ACS}=0.65$, and assuming 
 that $M_*\pm 0.5$~mags following \citealt{cross}).  
 The number density estimated from the high-redshift population is 
 smaller than that estimated from lower-redshifts.}
\label{fig:phiacs}
\end{figure}

In light of these arguments, it is interesting to consider Figures~2
and 4 in the main text.  If the Bernardi et al.\ sample does
underestimate the number density of early-type galaxies and we were
to increase the normalization $\phi_*$ to compensate, this would bring
the SDSS MVF closer to the SSRS2 IVF at lower $\sigma$.  However, the
large-$\sigma$ tails would still differ because the SDSS MVF correctly 
includes the effect of scatter in the $\sigma-L$ relation.  As for the
SDSS IVF, since it was obtained by transforming the luminosity function,
the final cut on signal-to-noise mainly affects the normalization but
does not substantially change the shape of the function.

On the other hand, the IVF is also sensitive to the relation between 
$\sigma$ and $L$:  in particular, it is sensitive to 
$\langle\sigma|L\rangle$.  So it is interesting to compare this
relation in the SDSS sample with results from the literature.  
Figure~\ref{fig:fj} shows this relation in the SDSS early-type sample; 
crosses show the median $V$ in small bins in $M_r$ where $M_r$ has
been corrected for evolution to $z=0$.  
The solid line shows a linear fit of the form given in the panel.  
For comparison, the dotted line shows the fit reported by 
\citet{bernardi}, 
 $\langle V|M_r\rangle = 2.2 - 0.105 (M_r + 21.15-0.125)$, 
where the factor of 0.125 accounts for the new SDSS photometry.  
The dots show a sample of 236 early-type galaxies from the compilation 
of \citet{prugniel}, rescaled to account for the shift in 
waveband (we used $M_B-M_r=1.32$ following Fukugita et al. 1995) and 
$H_0$ (they assumed $H_0=75$~km~s$^{-1}$~Mpc$^{-1}$, whereas we use 
70~km~s$^{-1}$~Mpc$^{-1}$).  
The dashed line shows $\langle V|M_r\rangle$; it is similar to the fit 
reported by \citet{forbes}:   
$V = 0.243 - 0.102 [M_r + 1.32 + 5{\rm Log}_{10}(75/70)]$.
As a result, the $\sigma-L$ relation has larger velocity dispersions 
for a given luminosity than the relation we use to estimate the 
IVF in the main text.  
The reason for this difference is clear:  inspection of the points 
suggests that the \citet{prugniel} sample has fewer luminous objects 
($M_r<-22.5$) with low $\sigma$ than one might have expected from 
inspection of the distribution of objects at lower luminosities in the 
sample.  The SDSS sample shows no such difference.  

\begin{figure}[t]
\begin{center}
\epsfig{file=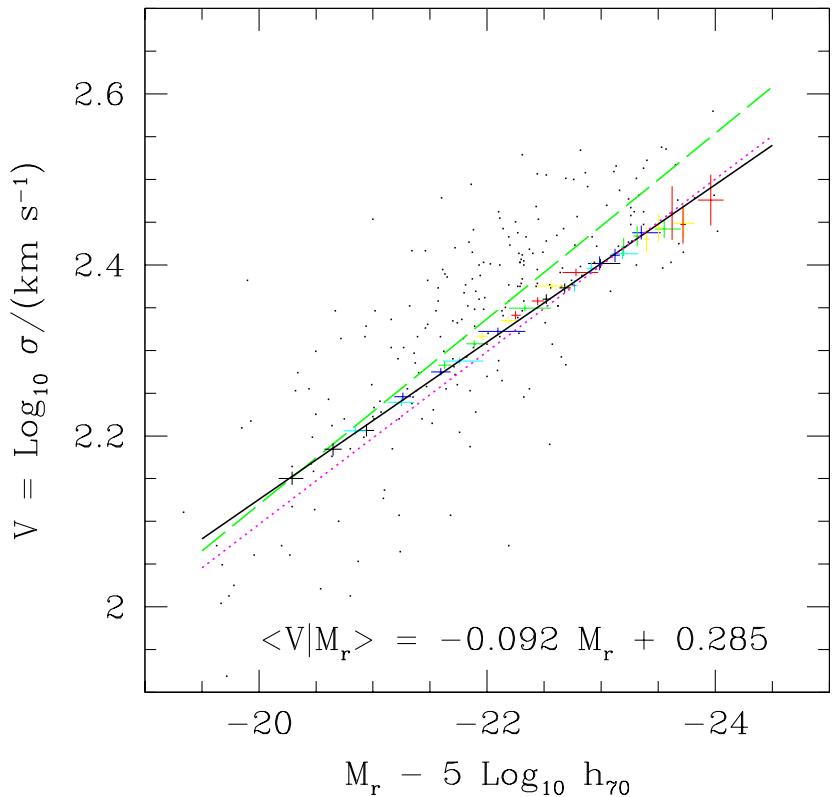,angle=0,width=4.7truein}
\end{center}
\caption{Correlation between velocity dispersion and luminosity.  
 Error-bars show the SDSS early-type galaxy sample, obtained by 
 stacking together many volume limited catalogs.  
 Solid line shows the fit reported in the figure, and dotted line 
 shows the fit from \citet{bernardi}, rescaled to account for 
 the new photometry.  Dots show a local sample of 236 early-type 
 galaxies from \citet{prugniel}, and dashed line shows a 
 fit to their sample.}
\label{fig:fj}
\end{figure}

\end{document}